\documentclass[11pt,twoside,draftcls,onecolumn]{IEEEtran}
\usepackage[utf8x]{inputenc}
\usepackage{amsfonts,amssymb,amsmath,amsthm,dsfont,bbm}
\usepackage{color}
\usepackage{algorithm,algorithmic}
\usepackage{graphicx}
\usepackage{subfig}
\usepackage{caption}
\newtheorem{lemma}{Lemma}
\newtheorem{result}{Proposition}
\newtheorem{theo}{Theorem}
\newtheorem{cor}{Corollary}

\theoremstyle{remark}

\newcommand\mI{\mathbf{I}}
\newcommand\mJ{\mathbf{J}}
\newcommand\mIJ{\left(\mI-\mJ\right)}
\newcommand\mIJkIJ{\left(\mIJ\kro\mIJ\right)}
\newcommand\mM{\mathbf{M}}
\newcommand\mK{\mathbf{K}}
\newcommand\mA{\mathbf{A}}
\newcommand\mB{\mathbf{B}}
\newcommand\mD{\mathbf{D}}
\newcommand\mR{\mathbf{R}}
\newcommand\mS{\mathbf{S}}
\newcommand\mQ{\mathbf{Q}}

\newcommand\mL{\mathbf{L}}
\newcommand\mP{\mathbf{P}}

\newcommand\mC{\mathbf{C}}
\newcommand\IDm{(\mI+\mD)^{-1}}

\newcommand\mk{m_{\mathcal{K}}}
\newcommand\pk{p_{\mathcal{K}}}

\newcommand\vOne{\mathbf{1}}
\newcommand\vx{\mathbf{x}}
\newcommand\vv{\mathbf{v}}
\newcommand\vs{\mathbf{s}}
\newcommand\vw{\mathbf{w}}
\newcommand\vu{\mathbf{u}}

\newcommand\kro{\otimes}

\newcommand\LRN{\left|\!\left|\!\left|}
\newcommand\RRN{\right|\!\right|\!\right|}

\newcommand\E{\mathbb{E}}
\newcommand\Prob{\mathbb{P}}
\newcommand{\e}{{\mathrm e}}
\newcommand\T{^{\mathrm T}}

\newcommand\ie{{\it i.e. }}
\newcommand\etal{{\it et al. }}

\title{Analysis of Sum-Weight-like algorithms for averaging in Wireless Sensor Networks}
\author{Franck~Iutzeler, Philippe~Ciblat and Walid~Hachem\thanks{The authors are with Institut Mines-T\'el\'ecom/T\'el\'ecom ParisTech ; CNRS LTCI. This work was partially granted by the French Defense Agency (DGA) and by the T\'el\'ecom/Eurecom Carnot Institute. Some parts of the work introduced  in this paper were presented at ICASSP 2012 \cite{Iutzeler2012}.}  } \date{}

\begin{document}
\maketitle

\markboth{submitted for publication to IEEE Transactions on Signal Processing, September 2012}{submitted for publication to IEEE Transactions on Signal Processing, September 2012}

\begin{abstract}
Distributed estimation of the average value over a Wireless Sensor Network has recently received a lot of attention. Most papers consider single variable  sensors and communications with feedback (e.g. peer-to-peer communications). However, in order to use efficiently the broadcast nature of the wireless channel, communications without feedback are advocated. To ensure the convergence in this feedback-free case, the recently-introduced Sum-Weight-like algorithms which rely on two variables at each sensor are a promising solution. In this paper, the convergence towards the consensus over the average of the initial values is analyzed in depth. Furthermore, it is shown that the squared error decreases exponentially with the time. In addition, a powerful algorithm relying on the Sum-Weight structure and taking into account the broadcast nature of the channel is proposed. 
\end{abstract}

%**************************************************************************%
%******************* Intro ************************************************%
%**************************************************************************%
\section{Introduction}\label{sec:intro}

The recent years have seen a surge of signal processing and estimation technologies operating in stressful environments. These environments do not make possible the use of a fusion center so the units/sensors have to behave in a distributed fashion. In various applications, sensors have to communicate through wireless channels because of the lack of infrastructure. Hence, along with distributed computation, the problem of communicating between the different sensors to estimate a global value is a key issue and was pioneered by Tsitsiklis \cite{Tsitsiklis1984}.

One of the most studied problems in Wireless Sensor Networks is the average computation of  the initial measurements of the sensors. More precisely, each sensor wants to reach \emph{consensus} over the mean of the initial values. A basic technique to address this problem, called \emph{Random Gossip}, is to make the sensors exchange their estimates in pairs and average them. This technique has been widely analyzed in terms of convergence and convergence speed in \cite{Boyd2004,Boyd2006,Dimakis2010}.

Finding more efficient exchange protocols has been a hot topic for the past few years; the proposed improvements were essentially twofold: i) exploiting the geometry of the network to have a more efficient mixing between the values (e.g. \cite{Dimakis2008,Benezit2010a,Ustebay2010}) and ii) taking advantage of the broadcast nature of the wireless channels (e.g. \cite{Aysal2009} without feedback link, and \cite{Nazer2009} with feedback link). Whereas the use of network geometry has received a lot of attention, the use of the broadcast nature of the wireless channel is less studied  albeit promising. Therefore, in our paper, we will focus on averaging algorithms taking into account the broadcast nature of the channel. In order to keep the number of communications as low as possible, we forbid the use of feedback links.

In the feedback-free context, one can mention \cite{Aysal2009}. However, even if the algorithm described in \cite{Aysal2009} converges quickly to a consensus, the reached value is incorrect. This can be explained by the fact that the sum of the sensor estimates is not constant over time. To overcome this problem, Franceschelli \etal \cite{Franceschelli2011} proposed to use well-chosen updates on two local variables per sensor while using the broadcast nature of the channel without feedback link. A more promising alternative is to use the \emph{Sum-Weight} scheme proposed by Kempe \cite{Kempe2003} and studied more generally by B\'en\'ezit \cite{Benezit2010}. In this setup, two local variables are also used: one representing the \emph{sum} of the received values and the other representing the \emph{weight} of the sensor (namely, the proportion of the sensor activity compared to the others). The two variables are transmitted at each iteration and both are updated in the same manner. The wanted estimate is then the quotient of these values. The convergence of this class of algorithms (without necessarily sum-conservation) has been proven in \cite{Kempe2003,Benezit2010}. In contrast, their convergence speed has never been theoretically evaluated except in \cite{Kempe2003} for a very specific case. 

The goal of this paper is to theoretically analyze the convergence speed for any Sum-Weight-like algorithm. As a by-product, we obtain necessary and sufficient condition for the convergence. In addition, we propose a new Sum-Weight-like algorithm based on broadcasting which outperforms existing ones. 

This paper is organized as follows: the notations and assumptions on the network model and on the Sum-Weight-like algorithms are provided in Section~\ref{sec:model}. Section~\ref{sec:math} is dedicated to the theoretical analysis of the squared error of the algorithms and provides the main contributions of the paper. In Section \ref{sec:algo}, we propose new Sum-Weight-like algorithms. In Section~\ref{sec:ana}, we compare our results with previous derivations done in the literature for the Sum-Weight-like algorithms as well as the algorithms based on the exchange of a single variable between the nodes. Section~\ref{sec:num} is devoted to numerical illustrations. Concluding remarks are drawn in Section~\ref{sec:cl}.

%**************************************************************************%
%******************* Model ************************************************%
%**************************************************************************%
\section{Model and Assumptions}\label{sec:model}

\subsection{Network model}

The sensor network will be modeled by a directed graph $\mathcal{G} = (V,E)$, $V$ being the set of vertices/sensors and $E$ being the set of edges which models the possible links between the sensors. We also define the adjacency matrix $\mA$ of $\mathcal{G}$ as the $N\times N$ matrix such that $\left(\mA\right)_{ij}$ equals $1$ if there is an edge from  $i$ to $j$ and $0$ otherwise. We define the neighborhood of each sensor $i$ as follows $\mathcal{N}_i = \left\lbrace j\in V \vert (i,j)\in E \right\rbrace$. Let $d_i= \vert \mathcal{N}_i  \vert$ denote the degree of the sensor $i$ where $\vert \mathcal{A} \vert$ represents the cardinality of the set $\mathcal{A}$. Let $d_{max}=\max_i d_i$ be the maximum degree. Let $\mD = \mathrm{diag}(d_1,\cdots,d_N)$ and  $\mL = \mD - \mA$ be the degree matrix and the Laplacian matrix respectively \cite{Biggs1993}.

Every sensor $i$ has an initial value $x_i(0)$ and we define $\vx(0) = [x_1(0),...,x_N(0)]^\mathrm{T}$ where the superscript~$^\mathrm{T}$ stands for the transposition. The goal of the network is to communicate through the edges of the underlying graph to reach consensus over the mean of the initial values of the sensors. A communication and estimation step will be referred to as an \emph{update}. 

We will assume that the network follows a discrete time model such that the time $t$ is the time of the $t$-th update. As an example, every sensor could be activated by an independent Poisson clock. The time would then be counted as the total number of clock ticks across the network. We will denote $x_i(t)$ the $i$-th sensor estimate at time $t$ and $\vx(t) = [x_1(t),...,x_N(t)]^\mathrm{T}$.

\subsection{Averaging Algorithms}

The goal of averaging algorithms is to make the vector of estimates $\vx(t)$ converge to $x_{ave}\vOne$, also known as the \emph{consensus vector}, where $\vOne$ is the length-$N$ vector of ones and $x_{ave} = (1/N) \vOne^\mathrm{T} \vx(0)$ is the average of the initial values of the sensors. In the present state-of-the-art, two classes of algorithms exist and are described below. 

\subsubsection{Class of Random Gossip algorithms}
\label{subsubsec:rg}

In standard gossip algorithms (e.g. \cite{Boyd2006}), sensors update their estimate according to the equation $\vx(t+1)^\mathrm{T} =  \vx(t)^\mathrm{T} \mK(t)$ where the $\mK(t)$ are doubly-stochastic matrices. We recall that a matrix $\mK$ is said row-stochastic (resp. column-stochastic) when all its elements are non-negative and when $\mK\vOne = \vOne$  (resp. $  \mK^\mathrm{T} \vOne = \vOne $). A matrix which is both row- and column-stochastic is said to be doubly-stochastic. Since two sensors can exchange information only across the edges of the graph, for any $ i \neq j$, $(\mK(t))_{ij}$ cannot be positive if $(\mA)_{ij}=0$.  From an algorithmic point of view, the row-stochasticity implies that the sum of the values is unchanged:  $\vx(t+1)^\mathrm{T} \vOne = \vx(t)^\mathrm{T} \mK(t) \vOne = \vx(t)^\mathrm{T}  \vOne  $ whereas the column-stochasticity implies that the consensus is stable: if $\vx(t) = c\vOne$, then $\vx(t+1)^\mathrm{T} = \vx(t)^\mathrm{T} \mK(t) = c \vOne^\mathrm{T} \mK(t) = c \vOne^\mathrm{T}$. For these reasons, double-stochasticity is desirable. However using doubly-stochastic matrices implies a feedback which is not always possible. In particular if a sensor sends information to multiple neighbors, the feedback message might raise multiple access problems. Similarly, if the message is sent through a long route within the network, the same route may not exist anymore for feedback in the context of mobile wireless networks. As these algorithms only rely on the exchanges of one variable per sensor, they will be called {\it single-variate}  algorithms in the rest of the paper.

\subsubsection{Class of Sum-Weight algorithms}
\label{subsubsec:sw}

To overcome this drawback, a possible method is to use two variables : one representing the \emph{sum} of the received values and another representing the relative \emph{weight} of the sensor. For the sensor $i$ at time $t$, they be respectively written $s_i(t)$ and $w_i(t)$. Writing $\vs(t) = [s_1(t),...,s_N(t)]^\mathrm{T}$ and $\vw(t) = [w_1(t),...,w_N(t)]^\mathrm{T}$, both variables will be modified by the same update matrix, $\vs(t+1)^\mathrm{T} =  \vs(t)^\mathrm{T} \mK(t)$ and $\vw(t+1)^\mathrm{T} =  \vw(t)^\mathrm{T} \mK(t)$. Finally, the estimate of sensor $i$ at time $t$ will be the quotient of the two variables, $x_i(t) \triangleq  s_i(t)/w_i(t)$. The initialization is done as follows:
\begin{equation}
\label{eq:init}
\left\{ \begin{array}{l}
 \mathbf{s}(0) =  \mathbf{x}(0) \\
 \mathbf{w}(0) =  \vOne .
\end{array}
\right.
\end{equation}

For the sake of convergence we will need an important property: \emph{ Mass Conservation} 
\begin{equation}
\label{eq:mass}
\left\{ \begin{array}{l}
		\sum_{i=1}^N s_i(t) = 	\sum_{i=1}^N x_i(0) = N x_{ave} \\ 
		\sum_{i=1}^N w_i(t) = N .
		\end{array}
		\right.
\end{equation}
This clearly rewrites as  $\forall t>0, \mK(t)\vOne = \vOne$ which corresponds to sum-conservation as in classic gossip algorithms and leads to row-stochastic updates matrices. 

\subsection{Notations for the Sum-Weight scheme}

Let us now introduce some useful notations along with some fundamental assumptions for convergence in the Sum-Weight scheme. Given two vectors $\mathbf{a}$ and $\mathbf{b}$ with the same size, we denote by $\mathbf{a}/\mathbf{b}$ the vector of the elementwise quotients. The Sum-Weight algorithm is described by the following equations: 
\begin{equation*} 
 \vx(t) \triangleq \frac{\vs(t)}{\vw(t)} = \left[ \frac{s_1(t)}{w_1(t)},...,  \frac{s_N(t)}{w_N(t)} \right]^{\mathrm{T}}
\end{equation*}
\begin{equation*}
\left\{ \begin{array}{l}
 \vs^\mathrm{T}(t+1) =  \vs^\mathrm{T}(t) \mK(t) = \vx^\mathrm{T}(0) \mP(t)   \\
 \vw^\mathrm{T}(t+1) =  \vw^\mathrm{T}(t) \mK(t) = \vOne^\mathrm{T} \mP(t)
\end{array}
\right.
\end{equation*}
with $\mP(t) = \mK(1)\mK(2)\dots\mK(t)$.

In the following, the matrix inequalities will be taken elementwise so that $\mM>0$ (resp. $\mM\geq0$) means that the matrix $\mM$ is (elementwise) \emph{positive} (resp. \emph{non-negative}). We recall that a non-negative matrix $\mM$ is said to be \emph{primitive} if $\mM^m > 0$ for some $m\geq1$ (see \cite[Chap~8.5]{Horn2005} for details).  We will denote the Kronecker product by `$\kro$'.

We can notice that reaching consensus is equivalent for $\vx(t)$ to converge to the consensus line $c\vOne$ where $c$ is consensus value. For this reason, it is useful to define $\mJ = (1/N) \vOne \vOne^\mathrm{T}$ the orthogonal projection matrix to the subspace spanned by $\vOne$ and $\mIJ$ the orthogonal projection matrix to the complementary subspace which can be seen as the error hyperplane. The matrix $\mathbf{I}$ is the identity matrix with appropriate size.

In order to intuitively understand the algorithm behavior,  let us decompose $\vx^\mathrm{T}(t)$ as follows 

\begin{eqnarray}
\nonumber \vx\T(t) &=& \frac{\vs\T(t)}{\vw\T(t)} =   \frac{\vx\T(0) \mP(t) }{\vw\T(t)} \\
\nonumber &=& \frac{\vx\T(0) \mJ \mP(t) }{\vw\T(t)} +  \frac{\vx\T(0) ( \mI - \mJ  ) \mP(t) }{\vw\T(t)} \\
\nonumber &=& \frac{x_{ave}\vOne\T  \mP(t) }{\vOne\T \mP(t)} +  \frac{\vx\T(0) ( \mI - \mJ  ) \mP(t) }{\vw\T(t)} \\
\label{eq:xinsight} &=& x_{ave} \vOne\T +  \frac{\vx\T(0) ( \mI - \mJ  ) \mP(t) }{\vw\T(t)} 
\end{eqnarray}
Obviously, the algorithm will converge to the right consensus if the second term in the right hand side vanishes. Actually, under some mild assumptions related to the connectedness of the network, we expect the numerator which corresponds to a projection on the error hyperplane will converge to zero at an exponential rate while all the elements of $\vw(t)$ are of order one. Proving these results will be the core of the paper. 

\subsection{Assumptions on the update matrices $\mathbf{K}(t)$}

First, we will always assume that both following conditions will be satisfied by any update matrix associated with a Sum-Weight like algorithm. 
\begin{description}
\item[{\bf(A1)}] Matrices $\left( \mK(t) \right)_{t>0} $ are independent and identically distributed (i.i.d.), and row-stochastic. The matrix $\mK(t) $ is valued in a set $\mathcal{K} = \left\{ \mK_i \right\}_{i=1..M}$ of size $M <  \infty$.  Also, $p_i = \Prob[\mK(t) = \mK_i ]>0$. 
\item[{\bf(A2)}] Any matrix in $\mathcal{K}$ has a strictly positive diagonal.
\end{description}
The first assumption is just a reformulation of the \emph{mass conservation} property introduced in section \ref{subsubsec:sw} along with the assumption of a finite number of actions across the network. This assumption is reasonable when one assumes that each sensor performs a finite number of actions. The second assumption forces every sensor to keep a part of the information it had previously. We also define 
\begin{eqnarray}
\label{eq:defm}
\left\{
\begin{array}{l}
\mk = \min_{i,j,k} \left\{ \left( \mK_k \right)_{ij} : \left( \mK_k \right)_{ij} > 0 \right\} , \\
\pk = \min_{k} \left\{ \Prob\left[ \mK(t) = \mK_k \right]  \right\} = \min_{k} p_k > 0  . \\
\end{array}
\right.
\end{eqnarray}

In addition to both previous assumptions, we will see that next assumption plays a central role in the convergence analysis of any Sum-Weight like algorithm. 
\begin{description}
\item[{\bf(B)}]  $ \E[\mK] = \sum_{i=1}^M p_i \mK_i $ is  a primitive matrix.
\end{description}

In terms of graph theory, matrix $\E[\mK]$ represents a weighted directed graph (see \cite[Def.~6.2.11]{Horn2005}). Since it is primitive, this graph is strongly connected (see \cite[Cor.~6.2.18]{Horn2005} and \cite{bullo2008}). Observe that this graph contains a self-loop at every node due to Assumption {\bf (A2)}. In fact, the matrix $\mA + \mI$ coincides with the so-called \emph{indicator matrix} (\cite[Def.~6.2.10]{Horn2005})  of $\E[\mK]$.

%**************************************************************************%
%******************* Maths ************************************************%
%**************************************************************************%
\section{Mathematical results}
\label{sec:math}

%**************************************************************************%
\subsection{Preliminary results}
The assumption {\bf (B)} can be re-written in different ways thanks to the next Lemma.

\begin{lemma} Under assumptions {\bf (A1)} and  {\bf (A2)}, the following propositions are equivalent to {\bf (B)} :
\label{lemma:eqB}
\begin{itemize}
\item[\bf (B1)] $\forall (i,j)\in \{1,...,N\}^2$, $\exists L_{ij}< N$  and a realization of $\mP(L_{ij})$ verifying $\mP(L_{ij})_{i,j}>0$. 
\item[\bf (B2)] $\exists L<2N^2$ and a realization of $\mP(L)$ which is a positive matrix.
\item[\bf (B3)] $\E[\mK\kro\mK] = \sum_{i=1}^M p_i \mK_i \kro \mK_i  $ is a primitive matrix.
\end{itemize}
\end{lemma}

The proof is reported in Appendix~\ref{anx:eq}. This Lemma will be very useful in the sequel since it enables us to interpret the Assumption {\bf (B)} in various manners.

Our approach for analyzing the convergence of Sum-Weight algorithms is inspired by \cite{Kempe2003} (with a number of important differences explained below) and so relies on the analysis of the Squared Error ($\mathrm{SE}$). Actually, the Squared Error can be upper-bounded by a product of two terms as follows
\begin{eqnarray}
\label{eq:sq} \|\mathbf{x}(t) - x_{ave}\mathbf{1} \|_2^2 &=& \sum_{i=1}^N \left|x_i(t) - x_{ave}\right|^2 = \sum_{i=1}^N \frac{1}{w_i(t)^2} \left|s_i(t) - x_{ave}w_i(t) \right|^2 \\
\nonumber &=& \sum_{i=1}^N \frac{1}{w_i(t)^2} \left| \sum_{j=1}^N x_j(0)\mathbf{P}_{ji}(t) - \frac{1}{N}\sum_{k=1}^N x_k(0).\sum_{l=1}^N \mathbf{P}_{li}(t)   \right|^2\\
\label{eq:mse2}
&\leq & \Psi_1(t) \Psi_2(t)
\end{eqnarray}

\begin{eqnarray}
\label{eq:psi1}
\textrm{with }~~~~~~~~~~~~~~~~~~ \Psi_1(t) &=&\frac{\|\vx(0)\|_2^2}{[\underset{k}{\min} \ w_k(t)]^2} \\
\label{eq:psi2}
\Psi_2(t) &=& \sum_{i=1}^N  \sum_{j=1}^N \left| \left( \mP^\mathrm{T}(t)\mIJ \right)_{ij} \right|^2.
\end{eqnarray}
Notice that  the decomposition done in Eq.~(\ref{eq:mse2}) mimics Eq.~(\ref{eq:xinsight}) for the Squared Error.

From now, our main contributions will be to understand the behavior of both terms $\Psi_1(t)$ and  $\Psi_2(t)$ when $t$ is large. In Section \ref{sec:psi1}, we will prove that $\Psi_1(t)$ can be upper bounded infinitely often. The term $\Psi_2(t)$ represents the projection of the current sensor values on the orthogonal space to the consensus line. The analysis of this term is drawn in Section \ref{sec:psi2}. 

%**************************************************************************%
\subsection{Analysis of  $\Psi_1(t)$}\label{sec:psi1} 

This term depends on the inverse of the minimum of the sensors weights (see Eq.~(\ref{eq:psi1})) and thus can increase quickly. However, the sensors frequently exchange information and hence spread their weight so the probability that a node weight keeps decreasing for a long time is very small. We will work on this probability and show that it can be made as small as one wants considering a sufficiently long amount of time. This will enable us to prove that $\Psi_1(t)$ will be infinitely often lower than a finite constant. To obtain these results, some preliminary lemmas are needed.

First, we will focus on the behavior of the nodes weights and especially on their minimum. One can remark that at every time $t$ there is as least one node whose weight is greater than or equal to $1$ (as the weights are non-negative and $\forall t>0, \sum_i w_i(t) = N$ because of the mass conservation exhibited in Eq.~(\ref{eq:mass})). As $\vw(t_0 + t)^\mathrm{T} = \vw(t)^\mathrm{T} \mP(t_0,t_0+t)$ where $\mP(t_0,t_0+t) \triangleq \mK(t_0)...\mK(t_0+t)$, it is interesting to focus on i) the minimum non-null value of $\mP(t_0,t_0+t)$ and ii) on the instants where $\mP(t_0,t_0+t)$ is positive.

\begin{lemma}
\label{lemma:P}
For all $t,t_0>0$, all the non-null coefficients of $\mP(t_0,t_0+t) $ are greater than or equal to $(\mk)^t$.
\end{lemma}

\begin{IEEEproof}
Let us recall that $\mk$ is the smallest non-null entry of all the matrices belonging to the set $\mathcal{K}$ as defined in Eq.~(\ref{eq:defm}). Let us consider the random matrix $\mP(t)$ (as the matrix choice is i.i.d., we drop the offset $t_0$). We will then prove this result by induction. It is trivial to see that every non-null coefficient of $\mP(1) = \mK(1)$ is greater than $\mk$ and  as
\begin{equation*}
\left( \mP(t) \right)_{i,j} = \sum_{k=1}^N  \left( \mP(t-1) \right)_{i,k} \left( \mK(t) \right)_{k,j},
\end{equation*}
it is obvious that if $\left( \mP(t) \right)_{i,j} > 0$, then there is a term in the sum that is positive (we remind that all the coefficient here are non-negative). This term is the product of a positive coefficient of $\mP(t-1) $ and a  positive coefficient of $\mK(t)$. Hence, if all the non-null coefficients of $\mP(t-1) $ are greater than $(\mk)^{t-1}$, then any non-null coefficient of $\mP(t)$ is greater than $(\mk)^{t-1}.\mk = (\mk)^t$. So, by induction, we have that $\forall t>0$ every non-null coefficient of $\mP(t) $ is greater than $(\mk)^t$. 
\end{IEEEproof}

Thanks to Item {\bf (B2)} of Lemma~\ref{lemma:eqB}, there is a finite $L$ such that there exists a realization of $\mP(L)$ which is a positive matrix. Considering the time at multiples of $L$, we know that for any $n$, if $\mP(nL+1,(n+1)L)>0$ then for all $i$, $w_i((n+1)L)\geq \mk ^L$. Let us define the following stopping times:
\begin{eqnarray}
\left\{
\begin{array}{l}
\nonumber \tau_0 = 0 \\
\tau_n = L \times \min \left\{ j : \sum_{k=1}^j \mathbbm{1}_{ \{ \mP(kL+1,(k+1)L) > 0  \}} = n   \right\} 
\end{array}
\right.
\end{eqnarray}
where $\mathbbm{1}_E$ is the indicator function of event $E$. And,
\begin{eqnarray}
\nonumber \Delta_n = \tau_n - \tau_{n-1} ~~~~~~ n=1,...,\infty.
\end{eqnarray}
The $\mathbbm{1}_{ \{ \mP(kL+1,(k+1)L) > 0  \}}$ are i.i.d. Bernoulli random variables with  strictly positive parameter $p$. Thus the inter-arrival times $\Delta_n$ are i.i.d. and geometrically distributed \ie $\Prob[\Delta_1=k]=p^{k-1}(1-p)$ for $k\geq1$. Observe that the $(\tau_n)_{n>0}$ are all finite and converge to infinity with probability one.  We then have proven the following result:
\begin{result}
\label{lemma:psi1}
Under Assumptions  {\bf (A1)}, {\bf (A2)}, and {\bf (B)}, there exists a sequence of positive i.i.d. geometrically distributed random variables $(\Delta_n)_{n>0}$ such that for all $n>0$ 
\begin{equation*}
 \Psi_1(\tau_n)\leq \|\vx(0)\|_2^2 (\mk)^{-2L}
\end{equation*}
where $\tau_n = \sum_{k=1}^n \Delta_k$.
\end{result}

%**************************************************************************%
\subsection{Analysis of  $\Psi_2(t)$}\label{sec:psi2} 

This section deals with new results about $\Psi_2(t)$. These results extend dramatically those given in \cite{Kempe2003} since we consider more general models for $\mathbf{K}(t)$ and any type of connected graph. According to Eq.~(\ref{eq:psi2}), we have 
\begin{equation} \label{eq:phi}
\Psi_2(t)= \| \mIJ \mP(t) \|_F^2
\end{equation}
where $\| \bullet \|_F$ denotes the Frobenius matrix norm.

One technique (used in {\it e.g.} \cite{Boyd2006}  ) consists in writing  $\mathbb{E}[\Psi_2(t)]=\mathrm{Trace} \left( (\mathbf{I}-\mathbf{J}) \mathbb{E} \left[\mathbf{P}(t) \mathbf{P}^\mathrm{T}(t) \right] (\mathbf{I}-\mathbf{J}) \right)$ thanks to Eq.~(\ref{eq:phi}) and finding a linear recursion between  $\mathbb{E}[\Psi_2(t) \vert \Psi_2(t-1)]$ and $\Psi_2(t-1)$. However this technique does not work in the most general case \footnote{We have  $\mathbb{E}[\Psi_2(t) \vert \Psi_2(t-1)]=\mathrm{Trace} \left( (\mathbf{I}-\mathbf{J}) \mP(t-1) \mIJ \mathbb{E}\left[\mathbf{K} \mathbf{K}^{T} \right] \mIJ \mP(t-1) (\mathbf{I}-\mathbf{J}) \right)$. By introducing   the matrix $ \mM = \mIJ \mathbb{E}\left[\mathbf{K} \mathbf{K}^{T} \right] \mIJ  $, it is easy to link $\mathbb{E}[\Psi_2(t) \vert \Psi_2(t-1)]$ with  $\Psi_2(t-1)$ since $\mathbb{E}[\Psi_2(t) \vert \Psi_2(t-1)] \leq \|\mM\|_{sp} \Psi_2(t-1) $ where $\|\bullet\|_{sp}$ is the spectral norm (see \cite[Chap.~7.7]{Horn2005} for details). Unfortunately, in some cases, $\|\mM\|_{sp}$ can be greater than 1; indeed for the \emph{BWGossip} algorithm  (introduced in Section~\ref{sec:BW}), one can have $\|\mM\|_{sp}>1$ for some underlying graphs. Nevertheless, this \emph{BWGossip} algorithm converges as we will see later. As a consequence, the inequality   $\mathbb{E}[\Psi_2(t) \vert \Psi_2(t-1)] \leq \|\mM\|_{sp} \Psi_2(t) $  is not tight enough to prove a general convergence result and another way has to be found.}.

Therefore, as proposed alternatively in \cite{Boyd2006} (though not essential in \cite{Boyd2006}) in the context of Random-Gossip Algorithms (see Section \ref{subsubsec:rg}), we write $\Psi_2(t)$ with respect to a more complicated matrix for which the recursion property is easier to analyze. Indeed, recalling that for any matrix $\mM$,
\begin{eqnarray*}
\| \mM \|_F^2  &=&  \mathrm{Trace}\left( \mM \mM\T  \right) \\
\textrm{and } ~~ \mathrm{Trace}\left( \mM \kro \mM   \right)  &=&  \left(\mathrm{Trace}\left( \mM  \right) \right)^2 
\end{eqnarray*} 
one can find that 
\begin{equation*}
\Psi_2(t) = \|\Xi(t)\|_F
\end{equation*}
with 
\begin{equation*}
 \Xi(t) = \mIJ\mP(t)\kro\mIJ\mP(t).
\end{equation*}
By remarking that $\mIJ\mP(t)\mIJ = \mIJ \mP(t)$, and by using standard properties on the Kronecker product, we have 
\begin{eqnarray}
\nonumber \Xi(t) &=& \mIJ\mP(t-1)\mIJ\mK(t)\kro\mIJ\mP(t-1)\mIJ\mK(t)  \\
&=&  \Xi(t-1)\left[ \left( \mIJ \kro \mIJ \right) \left(\mK(t)\kro \mK(t) \right) \right].
\label{eq:recurr}
\end{eqnarray}
By considering the mathematical expectation given the natural filtration of the past 
events $\mathcal{F}_{t-1} = \sigma\left( \mK(1),\cdots,\mK(t-1) \right)$, we obtain
\begin{equation*}
\E\left[\Xi(t) | \mathcal{F}_{t-1} \right] = \Xi(t-1) \left( \mIJ \kro \mIJ \right).\E\left[\mK \kro \mK \right] .
\end{equation*}
As $\Xi(0) = \mIJ \kro \mIJ$ and $\left(\mIJ \kro \mIJ\right)^2 = \mIJ \kro \mIJ $, we
finally have 
\begin{equation}
\label{eq:necsuff}
\E\left[\Xi(t) \right] = \mR^t.
\end{equation}
with 
\begin{equation}\label{eq:def_R}
\mR = \left( \mIJ \kro \mIJ \right).\E\left[\mK \kro \mK \right].
\end{equation}

 Now one can find a simple relationship between $\E[\Psi_2(t)]$ and the 
entries of the matrix $\E [ \Xi \left(t\right)]$ by considering $\mQ(t) = \mIJ \mP(t)$ and $\left( \mQ(t) \right)_{i,j} = q_{ij}(t)$. After simple algebraic manipulations, 
we show that 
$$\left( \E [ \Xi \left(t\right) ] \right)_{i+(k-1)N,j+(l-1)N} = \E [ q_{ij}(t) q_{kl}(t) ], \quad \forall (i,j,k,l) \in \{1,\cdots,N\}^4.$$
According to Eq.~(\ref{eq:phi}), we have $\E[\Psi_2(t)]= \E[ \| \mQ(t) \|_F^2 ]$
which implies that 
\begin{equation*}
\E[\Psi_2(t)] = \sum_{i,j=1}^N \E \left[ q_{ij}^2(t) \right]= \sum_{i,j=1}^N  \left( \E \left[ \Xi(t) \right] \right)_{i+(i-1)N,j+(j-1)N}. 
\end{equation*}
As a consequence, the behavior of the entries of $\E [ \Xi \left(t\right)]$ drives the behavior of $\E[\Psi_2(t)]$.

Let us define the $L_{\infty}$ vector norm on $N\times N$ matrices as $\LRN \mM \RRN_{\infty} = N \underset{1\leq i,j\leq N}{\max} |m_{ij}|$. The  norm $\LRN \bullet \RRN_{\infty}$ is a matrix norm (see \cite[Chap.~5.6]{Horn2005}) and hence is submultiplicative. Now, using the Jordan normal form of $\mR$ (see \cite[Chap.~3.1 and 3.2]{Horn2005}), we get that there is an invertible matrix $\mS$ such that
\begin{eqnarray}
 \label{eq:RJ}
\LRN \mR^t \RRN_{\infty} = \LRN \mS \mathbf{\Lambda}^t \mS^{-1} \RRN_{\infty} \leq \LRN \mS \RRN_{\infty} \LRN \mS^{-1} \RRN_{\infty} \LRN \mathbf{\Lambda}^t \RRN_{\infty}
\end{eqnarray}
where $\mathbf{\Lambda}$ is the Jordan matrix associated with $\mR$.

After some computations, it is easy to see  that the absolute value of all the entries of $\mathbf{\Lambda}^t$ are bounded in the following way:
\begin{eqnarray*}
\label{eq:maxJ}
 \underset{1\leq i,j\leq N}{\max} \left| (\mathbf{\Lambda}^t)_{ij}  \right| \leq \underset{ 0 \leq j \leq J-1}{\max} \binom{t}{t-j} \rho(\mR)^{t-j} 
\end{eqnarray*} 
with $\rho(\mR) $ the spectral radius of $\mR$ and $J$ the maximum size of the associated Jordan blocks. Hence, $\forall t>0$
\begin{equation}
\label{eq:maxJ2}
 \underset{1\leq i,j\leq N}{\max} \left| (\mathbf{\Lambda}^t)_{ij}  \right| \leq  t^{J-1} \rho(\mR)^{t-J+1} 
\end{equation} 
When  $\mR$ is diagonalizable, $J=1$, and we get that 
\begin{equation}\label{eq:maxJ3}
 \underset{1\leq i,j\leq N}{\max} \left| (\mathbf{\Lambda}^t)_{ij}  \right| \leq  \rho(\mR)^t \quad \textrm{(when $ \mR$ is diagonalizable)}
\end{equation} 

Putting together Eqs.~(\ref{eq:necsuff}), (\ref{eq:RJ}), (\ref{eq:maxJ2}), (\ref{eq:maxJ3}), and remarking that the subspace spanned by $\vOne_{N^2} = \vOne\kro\vOne$ is in the kernel of $\mR$, we get that the size of the greatest Jordan block is $\leq N-1$, hence the following lemma:

\begin{lemma}\label{lem:R} 
We have 
$$\E[\Psi_2(t)] = \mathcal{O}\left( t^{N-2} \rho(\mR)^t \right)$$
where $\mR$ is defined in Eq.~(\ref{eq:def_R}) and where $\rho(\mR)$ is the spectral radius of the matrix $\mR$.
\end{lemma}

The next step of our analysis is to prove that the spectral radius $\rho(\mathbf{R})$
is strictly less than $1$ when Assumptions {\bf (A1)}, {\bf (A2)}, and {\bf (B)} hold. 
Applying  Theorem~5.6.12 of \cite{Horn2005} on Eq.~(\ref{eq:necsuff}) proves that 
$\rho\left( \mR \right)<1$ if and only if $\E\left[\Xi(t) \right]$ converges to zero as $t$ goes to infinity. Therefore our next objective  is to prove that $\E\left[\Xi(t) \right]$ converges to zero by using another way than the study of the spectral radius of $\mR$. 

Actually, one can find another linear recursion on $\Xi(t)$ (different from the one exhibited in Eq.~(\ref{eq:recurr})). We get
$$\Xi(t)=\Xi(t-1).\left(\mK(t)\kro\mK(t) \right)$$
and, by taking the mathematical expectation given the past, we obtain
$$\E\left[\Xi(t) \vert \mathcal{F}_{t-1} \right]= \Xi(t-1).\E\left[\mK \kro \mK \right].$$
Remarking that $\Xi(t)\vOne_{N^2}=0$, we have for any vector $\vv$, 
$$\E\left[\Xi(t) \vert \mathcal{F}_{t-1} \right] = \Xi(t-1).\left(\E\left[\mK \kro \mK \right] - \vOne_{N^2} \vv^\mathrm{T}  \right)$$
and then, for any vector $\vv$,  
\begin{equation}\label{eq:nec} 
\E\left[\Xi(t) \right]=\Xi(0) \mS_{\vv}^t
\end{equation}
with $\mS_{\vv}= \E\left[\mK \kro \mK \right] -  \vOne_{N^2}\vv^\mathrm{T} $ and $\Xi(0) = \mIJ \kro \mIJ$. 

By considering Eq.~(\ref{eq:nec}), it is straightforward to see that
$\E\left[\Xi(t) \right]$ converges to zero as $t$ goes to infinity if 
there is a vector $\vv$ such that $\rho(\mS_{\vv})<1$. Notice that the recursion given in 
Eq.~(\ref{eq:nec}) is less ``strong'' than the one in Eq.~(\ref{eq:necsuff}) since it only leads 
to a sufficient condition instead of a necessary and sufficient condition. 
As $\rho(\mS_{\vv})<1$ implies the convergence of $\E\left[\Xi(t) \right]$ and as the convergence of $\E\left[\Xi(t) \right]$ implies that $\rho(\mR)<1$, one thus can state the following Lemma:

\begin{lemma}\label{lem:Rbis}
If there is a vector $\vv$ such that $\rho\left( \E\left[\mK \kro \mK \right] -  \vOne_{N^2}\vv^\mathrm{T}  \right)<1$, then $\rho(\mR) <1$ . 
\end{lemma}
One of the most important result in the paper lies in the following Lemma in which 
we ensure that, under Assumptions  {\bf (A1)}, {\bf (A2)}, and {\bf (B)} 
there is a vector $\vv$ such that $\rho\left( \E\left[\mK \kro \mK \right] -  \vOne_{N^2} \vv^\mathrm{T}  \right)<1$ and thus $\rho(\mR)<1$.
\begin{lemma}
\label{lemma:rho} If Assumptions {\bf (A1)}, {\bf (A2)}, {\bf (B)} hold,
there is a vector $\vv$ such that $\rho\left( \E\left[\mK \kro \mK \right] -  \vOne_{N^2} \vv^\mathrm{T}  \right)<1$.
\end{lemma}

\begin{IEEEproof}
Assumptions {\bf (A1)}, {\bf (A2)}, and {\bf (B)} imply that 
\begin{itemize}
\item[i)] $\E[\mK\kro\mK]$ is a non-negative matrix with a constant row sum equal to one (because of the row-stochasticity). According to Lemma~8.1.21 in \cite{Horn2005}, we have 
$\rho(\E[\mK\kro\mK]) = 1 $. 
\item[ii)] $\E[\mK\kro\mK]$ is a primitive matrix (see {\bf(B3)} in Lemma~\ref{lemma:eqB})
which implies that there only is one eigenvalue of maximum modulus. This eigenvalue is thus equal to $1$ and associated with the eigenvector $\vOne_{N^2}$.
\end{itemize}

By using the Jordan normal form and the simple multiplicity of the maximum eigenvalue (equal to $1$), we know that i) there exists a vector $\vv_1$ equal to the left eigenvector corresponding to the eigenvalue $1$, and ii) that the set of the eigenvalues of $\E\left[\mK \kro \mK \right]- \vOne_{N^2} \vv_1^\mathrm{T} = \mS_{\vv_1} $  are exactly the set of the eigenvalues of  $\E\left[\mK \kro \mK \right]$ without the maximum one equal to $1$. Indeed the maximum eigenvalue of $\E\left[\mK \kro \mK \right]$ has been removed by the vector $\vOne_{N^2} \vv_1^\mathrm{T}$ and the associated eigenvector now belongs to the kernel of $\mS_{\vv_1}$. As a consequence, the modulus of the eigenvalues of $\mS_{\vv_1}$ is strictly less than $1$, {\it i.e.}, $\rho(\mS_{\vv_1})<1$.
\end{IEEEproof}

Aggregating successively the results provided in Lemmas \ref{lemma:rho}, \ref{lem:Rbis}, and \ref{lem:R} leads to the main result of this Section devoted to the analysis of $\Psi_2(t)$. Indeed, Lemma \ref{lemma:rho} ensures that there is a vector $\vv$ such that $\rho(\mS_{\vv} )<1$, then Lemma \ref{lem:Rbis} states that $\rho(\mR )<1$. Then, Lemma \ref{lem:R} concludes the proof for the next result. 
\begin{result}
\label{lemma:psi2}
Under Assumptions  {\bf (A1)}, {\bf (A2)} and {\bf (B)} holds, then  
$$ \E[\Psi_2(t)] = \mathcal{O} \left( t^{N-2} e^{-\kappa t} \right)$$
with $\kappa  = - \log\left( \rho\left(\mR\right) \right) >0$.
\end{result}

%**************************************************************************%
\subsection{Final results}

Thanks to the various intermediate Lemmas and Propositions provided above, we are now able to state the main Theorems of the paper. The first one deals with the determination of the necessary and sufficient conditions for Sum-Weight-like algorithms to converge. The second one gives us an insight on the decrease speed of the Squared Error (defined in Eq.~(\ref{eq:sq})). In the meanwhile, we need the following lemma:
\begin{lemma}
\label{lemma:contr}
$\|\vx(t)-x_{ave}\vOne\|_\infty=\max_{i} | x_i(t)-x_{ave}|$ is a  non-increasing sequence with respect to $t$.
\end{lemma}

\begin{IEEEproof}
One can remark that, at time $t+1$, we have 
\begin{eqnarray*}
\forall j, \   x_j(t+1) &=& \frac{\sum_{i=1}^N  (\mK)_{ij} s_i(t)  }{\sum_{i=1}^N   (\mK)_{ij} w_i(t) }  \\
&=& \sum_{i=1}^N  \left( \frac{(\mK)_{ij} w_i(t)   }{\sum_{\ell=1}^N   (\mK)_{\ell j} w_\ell(t)}\right) x_i(t)
\end{eqnarray*}
where $\mK$ corresponds to any matrix in $\mathcal{K}$. So $x_j(t+1)$ is a center of mass of $(x_i(t))_{i=1,...,N}$. Therefore, $\forall j\in\{1,...,N\}$,
\begin{eqnarray*}
|x_j(t+1)-x_{ave}|  &\leq&  \sum_{i=1}^N  \left( \frac{(\mK)_{ij} w_i(t)   }{\sum_{\ell=1}^N   (\mK)_{\ell j} w_\ell(t)}\right) |x_i(t)-x_{ave}|\\
&\leq &  \max_i |x_i(t)-x_{ave}|.
\end{eqnarray*}
\end{IEEEproof}

%****************************************%
\subsubsection{Result on the convergence} 
\label{sec:resconv}
Let us consider that Assumption {\bf (B)} does not hold. Thanks to {\bf (B1)} in Lemma~\ref{lemma:eqB},  
this is equivalent to $\exists (k,l)\in N^2$ such that $ \forall T, \ \mP(T)_{k,l} = 0$. Let us take $\vx(0)$ equal to the canonical vector composed by a $1$ at the $k$-th position and $0$  elsewhere. Then for any $t>0$, $x_l(t) = 0$ which is different from $x_{ave}=1/N$. 
Consequently, the algorithm does not converge to the true consensus for any initial measurement. So if the Sum-Weight algorithm converges almost surely to the true consensus for any initial vector $\vx(0)$ then Assumption {\bf (B)} holds.

Let us now assume that Assumption {\bf (B)} holds. Using Markov's inequality along with Result~\ref{lemma:psi2}, we have a finite $K$ such that for any $\delta>0$,  
\begin{eqnarray*}
\sum_{t>0} \Prob\left[ \vert \Psi_2(t) \vert > \delta \right]&\leq&  \frac{1}{\delta} \sum_{t>0}   \E[ \vert \Psi_2(t) \vert] \\
&\leq&  \frac{1}{\delta} K  \sum_{t>0} t^{N-2} \e^{-\kappa t} < \infty .
\end{eqnarray*}

Consequently, Borel-Cantelli's Lemma leads to the almost sure convergence of $\Psi_2(t)$ to zero. In addition, the random variables $(\tau_n)_{n>0}$ provided in the statement of Proposition~\ref{lemma:psi1} converge to infinity with probability one, hence $\Psi_2(\tau_n) \to 0$ almost surely. Since $ \Psi_1(\tau_n) $ is bounded,  $\Psi_1(\tau_n)\Psi_2(\tau_n) \underset{n\to \infty}{\rightarrow} 0$ almost surely. According to Lemma~\ref{lemma:contr}, $\|\vx(t)-x_{ave}\vOne\|_\infty$ is a nonincreasing nonnegative sequence verifying $\|\vx(t)-x_{ave}\vOne\|_\infty \leq \Psi_1(t)\Psi_2(t)$, as there is converging subsequence with limit $0$, the sequence itself converges to the same limit which implies the following theorem.

\begin{theo}
\label{theo:main}
Under Assumptions  {\bf (A1)} and {\bf (A2)},  $\vx(t)$ converges almost surely to the average consensus $x_{ave}\vOne$ for any $\mathbf{x}(0)$, if and only if Assumption {\bf (B)} holds.
\end{theo}

We have additional result on another type of convergence for $\vx(t)$. As $\|\vx(t)-x_{ave}\vOne\|_\infty$ is a non-increasing sequence, we have, for any $t$,  $\|\vx(t)-x_{ave}\vOne\|_\infty \leq \|\vx(0)-x_{ave}\vOne\|_\infty$ which implies that $\vx(t)$ is bounded for any $t>0$. As a consequence, according to \cite{VanderVaart2000}, since $\vx(t)$ also converges almost surely to $x_{ave}\vOne$, we know that $\vx(t)$ converges to $x_{ave}\vOne$ in $L^p$ for any positive integer $p$. The convergence of the mean squared error of $\vx(t)$ thus corresponds to the case $p=2$. 
\begin{cor}
\label{cor:main}
If $\vx(t)$ converges almost surely to the average consensus $x_{ave}\vOne$ then the mean squared error (MSE) converges to zero.
\end{cor}

%**********************************************%
\subsubsection{Result on the convergence speed}

The next result on the convergence speed corresponds to the main challenge and novelty of the paper. Except in \cite{Kempe2003} for a very specific case (cf. Section \ref{sec:kempe} for more details), our paper provides the first general results about the theoretical convergence speed for the squared error of the Sum-Weight like algorithms. For the sake of this theorem we introduce the following notation: given two sequences of random variables $(X_n)_{n>0}$ and $(Y_n)_{n>0}$, we will say that $X_n = o_{\textrm{a.s.}}(Y_n)$ if $X_n / Y_n \to 0  $ almost surely.

\begin{theo}
\label{theo:speed}
Under Assumptions  {\bf (A1)}, {\bf (A2)}, and {\bf (B)}, the squared error ($\mathrm{SE}$) is non-increasing. Furthermore, it is bounded by an exponentially decreasing function as follows
\begin{eqnarray*}
 \mathrm{SE}(\tau_n) =  \| \vx(\tau_n) - x_{ave}\vOne \|_2^2   = o_{\textrm{a.s.}}\left( \tau_n^N \e^{-\kappa \tau_n}\right)
\end{eqnarray*}
with $\kappa = - \log \left( \rho\left(  \left(\mIJ \kro \mIJ\right) \E\left[\mK \kro \mK \right] \right) \right)>0$
and   $\tau_n = \sum_{i=1}^n \Delta_i $ as defined in Proposition~\ref{lemma:psi1}.
\end{theo}
This result tells us that the slope of $\log(\mathrm{SE(t)})$ is lower-bounded by $\kappa$ infinitely often which provides us a good insight about the asymptotic behavior of $\vx(t)$. Indeed, the squared error will vanish exponentially and we have derived a lower bound for this speed. We believe this result is new as it may foretell any algorithm speed. The particular behavior of the weights variables in this very general setting does not enable us to provide a clearer result about the mean squared error; however for some particular algorithms (e.g. single-variate ones) this derivation is possible (see Section~\ref{sec:ana} for more details). The authors would like to draw the reader's attention to the fact that the main contribution of the paper lies in the exponential decrease constant $\kappa$. %% exhibited in our numerical results (see Section~\ref{sec:num}).
\begin{IEEEproof}
To prove this result we will once more use the decomposition of the squared error introduced in Eq.~(\ref{eq:mse2}). We know from Proposition~\ref{lemma:psi2} that $\E[t^{-N} \e^{\kappa  t} \Psi_2(t)] = \mathcal{O}(t^{-2})$. By Markov's inequality and Borel-Cantelli's lemma, 
\begin{eqnarray*}
t^{-N} \e^{\kappa t} \Psi_2(t) \xrightarrow[t\to\infty]{} 0 ~~~\textrm{almost surely.}
\end{eqnarray*}
Composing with the $(\tau_n)_{n>0}$, we get
\begin{eqnarray*}
\tau_n^{-N} \e^{\kappa \tau_n} \Psi_2(\tau_n) \xrightarrow[n\to\infty]{} 0 ~~~\textrm{almost surely.}
\end{eqnarray*}
Since $\exists C, \forall n>0, \Psi_1(\tau_n)\leq C$, we get the claimed result.
\end{IEEEproof}

%**************************************************************************%
%******************* Algos ******************************%
%**************************************************************************%
\section{Proposed algorithms}\label{sec:algo}

In Subsection \ref{sec:BW}, we propose a new Sum-Weight-like algorithm using the broadcast nature of the wireless channel which converges and offers remarkable performance. This algorithm is hereafter called {\it Broadcast-Weighted Gossip (BWGossip)}. In Subsection \ref{sec:clock}, a new distributed management of the nodes' clocks which can improve averaging algorithms is proposed. Finally, Subsection \ref{sec:sum} provides an extension of this work to the distributed sum computation.

%**************************************************************************%
\subsection{BWGossip algorithm}\label{sec:BW}

Remarking i) that the broadcast nature of the wireless channel was often not taken into account in the distributed estimation algorithms (apart in \cite{Aysal2009} but this algorithm does not converge to the average) and ii) that information propagation is much faster while broadcasting compared to pairwise exchanges \cite{Iutzeler2012a}, we propose an algorithm taking into account the broadcast nature of the wireless channel. At each global clock tick, it simply consists in uniformly choosing a sensor that broadcasts its pair of values in an appropriate way; then, the receiving sensors add their received pair of values to their current one. A more algorithmic formulation is presented below.
\begin{algorithm}[h!]
\caption{BWGossip}
\label{algo:bwg}
\begin{algorithmic} 
\STATE When the sensor $i$ wakes up (at global time $t$):

\STATE $~~~\blacktriangleright $  The sensor $i$ broadcasts $\left(\frac{s_i(t)}{\vert\mathcal{N}_i\vert+1};\frac{w_i(t)}{\vert\mathcal{N}_i\vert+1}\right)$ \\

\STATE $~~~\blacktriangleright $ The sensors of the neighborhood $\mathcal{N}_i$ update : 
$\forall j\in\mathcal{N}_i,\ \left\{ \begin{array}{l}
 s_j(t+1) = s_j(t) + \frac{s_i(t)}{\vert\mathcal{N}_i\vert+1}  \\
 w_j(t+1) = w_j(t) + \frac{w_i(t)}{\vert\mathcal{N}_i\vert+1}
\end{array}
\right.$

\STATE $~~~\blacktriangleright $  The sensor $i$ updates : $
\left\{ \begin{array}{l}
 s_i(t+1) = \frac{s_i(t)}{\vert\mathcal{N}_i\vert+1}\\
 w_i(t+1) = \frac{w_i(t)}{\vert\mathcal{N}_i\vert+1}
\end{array}
\right.$

\end{algorithmic} 
\end{algorithm}

According to this formulation, the update matrix $\mK_i$ associated with the action of the $i$-th sensor takes the following form
\begin{eqnarray}
\nonumber \mK_i &=& \mI - e_i e_i^\mathrm{T}+ e_i e_i^\mathrm{T} \left[ \IDm \left( \mA +  \mI \right) \right] \\
&=& \mI - e_i e_i^\mathrm{T}  \IDm \mL \label{eq:Ki}
\end{eqnarray}
with $e_i$ the $i$-th canonical vector. Clearly, the update matrices satisfy the Assumptions {\bf(A1)} and {\bf(A2)}. 

Thanks to Eq.~(\ref{eq:Ki}) and recalling that $\mL=\mD-\mA$, we obtain that 
\begin{eqnarray*}
\E[\mK] &= &\mI - \frac{1}{N}\IDm \mL\\
 &=& \frac{N-1}{N}\mI + \IDm \left( \mI +  \mA \right).
\end{eqnarray*}
As all the involved matrices are non-negative, we have $\IDm \left( \mI +  \mA \right) \geq \left( \mI +  \mA \right)/((d_{max}+1)N)$. As a consequence, we have 
$$\E[\mK]\geq \frac{1}{(d_{max}+1)N} (\mI+\mA)\geq 0.$$ 
Since $\mA$ is the adjacency matrix of a connected graph, $\exists m>0, (\mI+\mA)^m > 0$. Hence, for the same $m$,  $\E[\mK]^m \geq  1/(d_{max}N+N)^m (\mI+\mA)^m   > 0$, which implies that $\E[\mK]$ is a primitive matrix. Applying Lemma~\ref{lemma:eqB} enables us to prove that Assumption  {\bf(B)} also holds. 

Hence, Theorem~\ref{theo:main} states that the BWGossip algorithm converges almost surely to the average consensus and Theorem~\ref{theo:speed} gives us an insight about the decrease speed of the squared error.

%**************************************************************************%
\subsection{Adaptation to smart clock management}
\label{sec:clock}

 So far, all the Poisson coefficients of the clocks were identical. This means that all sensors were waking up uniformly and independently from their past actions. Intuitively, it would be more logical that a sensor talking a lot became less active during a long period.

Another advantage of the Sum-Weight algorithms is the knowledge of how much a sensor talks compared to the others which is a useful information. Actually, each sensor knows whether it talks frequently or not (without additional cost) through its own weight value because when a sensor talks, its weight decreases and conversely when it receives information, its weight increases. Therefore, our idea is to control the Poisson coefficient of each sensor with respect to its weight.

We thus propose to consider the following rule for each Poisson coefficient
\begin{equation}
\label{eq:lambda}
\forall i\in V, ~~~ \lambda_i(t) = \alpha + (1-\alpha) w_i(t)  
\end{equation}
where $\alpha \in (0,1)$ is a tuning coefficient. 

Notice that the global clock remains unchanged since $\forall t>0, \  \sum_{i=1}^N \lambda_i(t) = N $. Keeping the global message exchange rate unchanged, the clock rates of each sensor are improved. The complexity of the algorithm is the same because the sensor whose weight changes has just to launch a Poisson clock.

\begin{figure}[ht]
\centering
\includegraphics[width = 0.9\columnwidth]{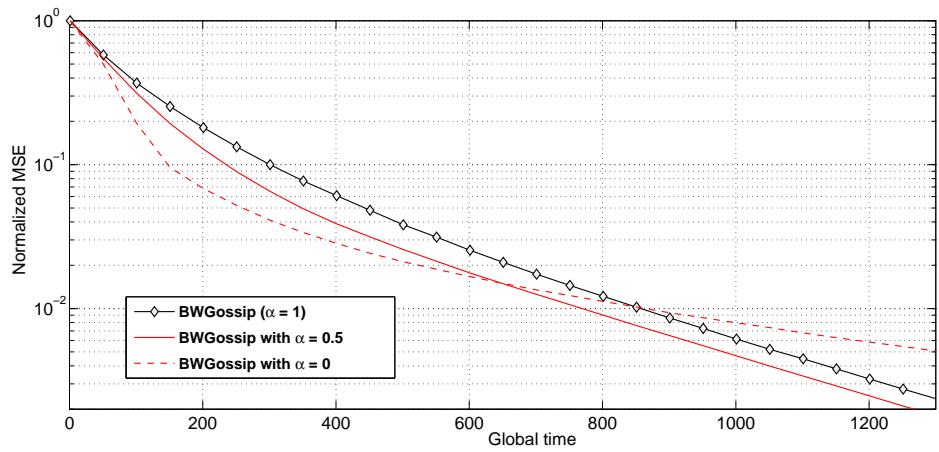}
\caption{\label{fig:clock} Mean squared error of the \emph{BWGossip} versus time for different clock management schemes.}
\end{figure}

Even if the convergence and the convergence speed  with clock improvement have not been formally established, our simulations with the BWGossip algorithm (see Fig.~\ref{fig:clock}) show that it seems to also converge exponentially to the average more quickly if $\alpha$ is well chosen.

%**************************************************************************%
\subsection{Distributed estimation of the sum}
\label{sec:sum}

In some cases, distributively computing the sum of the initial values is very interesting. For example, in the case of signal detection, the Log Likelihood Ratio (LLR) of a set of sensors is separable into the sum of the LLRs of the sensors. Hence, in order to perform a signal detection test based on the information of the whole network (using a Generalized LLR Test for instance), every sensor needs to estimate the sum of the LLRs computed by the sensors.

An estimate of the sum can be trivially obtained by multiplying the average estimate by the number of sensors which might not be available at any sensor. Another interest of the Sum-Weight scheme is that the initialization of the weights of the sensors enables us to compute different functions related to the average. Intuitively, as the sum of the $\vs(t)$ and $\vw(t)$ vectors are conserved through time and the convergence to a consensus is guaranteed by the assumptions on the update matrices, we get that the sensors will converge to $\sum_i s_i(0) / \sum_i w_i(0)$. This is obviously equal to the average $1/N \sum_i x_i(0)$ with the initialisation of Eq.~(\ref{eq:init}). 

Now, if a sensor wants to trigger a estimation of the sum through the network, it simply sets its weight to $1$ and sends a starting signal to the other nodes which set their weights to $0$. Mathematically, we then have the following initialization after sensor $i$ triggers the algorithm
\begin{equation*}
\left\{ \begin{array}{l}
 \mathbf{s}(0) =  \mathbf{x}(0) \\
 \mathbf{w}(0) =  e_i 
\end{array}
\right.
\end{equation*}
where $e_i$ is the $i$-th canonical vector. In this setting, all Sum-Weight like algorithms converge exponentially to the sum of the initial value as all the theorems of the paper hold with only minor modifications in the proofs.

%**************************************************************************%
%******************* State of the Art ******************************%
%**************************************************************************%
\section{Comparison with existing works}\label{sec:ana}

In this section, we will show that our results extend the works done previously in the literature. In Subsection \ref{sec:kempe} and \ref{sec:bene}, we compare our results with existing papers dealing with the design and the analysis of the Sum-Weight like algorithms. In Subsection \ref{sec:mono}, we will observe that our results can even be applied to the traditional framework of single-variate  gossip algorithms.

%**************************************************************************%
\subsection{Comparison with Kempe's work} \label{sec:kempe}

In the Kempe's work  \cite{Kempe2003}, the setup is quite different since the sensors' updates are synchronous, that is, at each time $t$, all the sensors send and update their values. Another important difference lies in the fact that the communication graph is assumed to be complete and to offer self-loops, i.e.,  each sensor can communicate with any other one, including itself. The algorithm introduced in \cite{Kempe2003} is described in Algorithm \ref{algo:kempe}. 
\begin{algorithm}[h!]
\caption{Push-Sum Algorithm  \cite{Kempe2003}}
\label{algo:kempe}
\begin{algorithmic} 
\STATE At each time $t$, every sensor $i$ activates:

\STATE $~~~\blacktriangleright $  The sensor $i$ chooses uniformly a node $j_i(t)$ belonging to its neighborhood (including itself)\\

\STATE $~~~\blacktriangleright $ The sensor $i$ sends the pair $(s_i(t)/2;w_i(t)/2)$ to $j_i(t)$\\

\STATE $~~~\blacktriangleright $  Let $\mathcal{R}$ be the set of sensors that sent information to $i$. 
The sensor $i$ updates:  
$$\left\{ \begin{array}{l}
 s_i(t+1) = s_i(t)/2 + \sum_{r\in \mathcal{R}}  s_r(t)/2\\
 w_i(t+1) = w_i(t)/2+ \sum_{r\in \mathcal{R}} w_r(t)/2
\end{array}
\right.$$
\end{algorithmic} 
\end{algorithm}

Consequently, at time $t$, the update matrix takes the following form
\begin{equation}\label{eq:kk}
\mK(t) = \frac{1}{2} \mI + \frac{1}{2} \sum_{i=1}^N e_i e_{j_i(t)}^\mathrm{T}
\end{equation}
where the index $j_i(t)$ is defined in Algorithm \ref{algo:kempe}. 
Notice that the first term of the right hand side corresponds to the information kept by the sensor, while the second term corresponds to 
the information sent to the chosen sensor. Moreover, as each sensor selects uniformly its neighbor\footnote{as the graph is complete, this means, choosing one node uniformly in the graph.}
 (including itself), we obtain that
\begin{eqnarray*}
\E[\mK] &=& \frac{1}{2} \mI + \frac{1}{2} \mJ.
\end{eqnarray*}
It is then easy to check that 
\begin{itemize}
\item[-] the (instantaneous) update matrices are non-negative and row-stochastic. In addition, they are chosen uniformly in a set of size $N^N$. 
\item[-] the (instantaneous) update matrices have a strictly positive diagonal.  
\item[-] $\E[\mK]>0$, thus $\E[\mK]$ is a primitive matrix.
\end{itemize}
This proves that the Kempe's algorithm satisfies the assumptions {\bf(A1)}, {\bf(A2)} and {\bf(B)}, and so  it converges almost surely to the average consensus (which was also proven in \cite{Kempe2003}). 

Let us now focus on the convergence speed of the Kempe's algorithm. 
We remind that the convergence speed is driven by $\Psi_2(t)$ (denoted by $\Phi_t$ in \cite{Kempe2003}). As this algorithm is synchronous and only applies on a complete communication graph, it is simple to obtain a  recursion between $\E[\Psi_2(t)\vert\Psi_2(t-1)]$ and $\Psi_2(t-1)$. Indeed, the approach given in the footnote of Section~\ref{sec:psi2} can be applied. More precisely, 
the corresponding matrix $\mM = \mIJ \E[\mK\mK^\mathrm{T}] \mIJ$ is given in closed-form as (see Appendix~\ref{anx:kempeL2} for details) 
\begin{equation}\label{eq:speed_kempe}
\mM = \mIJ \E[\mK\mK^\mathrm{T}] \mIJ =   \left( \frac{1}{2}  -  \frac{1}{4N} \right)\mIJ , 
\end{equation}
and then one can easily check\,\footnote{Note that there is a typo in Lemma 2.3 of \cite{Kempe2003}. Indeed, the coefficient is $(1/2 - 1/(2N))$ in \cite{Kempe2003} instead of $(1/2-1/(4N))$.} that 
\begin{equation}\label{eq:rec_kempe}
 \E[\Psi_2(t)\vert\Psi_2(t-1)] = \left( \frac{1}{2} - \frac{1}{4N} \right) \Psi_2(t-1). 
\end{equation}
Moreover, thanks to Eq.~(\ref{eq:speed_kempe}), we have that $\rho(\mM)=\left(1/2  -  1/(4N) \right) < 1$ and thus the inequality in the above-mentioned footnote has been replaced with an equality and the spectral radius of $\mM$ is less than $1$. 
Therefore, the  true convergence speed is provided by $\rho(\mM)$. Comparing this previous convergence speed (obtained very easily in  \cite{Kempe2003}) with the convergence speed bounds obtained in our paper is of great interest and will be done below. 
First of all we remind (see the footnote  in Section~\ref{sec:psi2}) that in the general case treated in our paper, it is impossible to find a recursion similar to Eq.~(\ref{eq:rec_kempe}) which justifies our alternative approach. Secondly, following the general alternative approach developed in this paper, we know that the matrix of interest is 
$\mR = \mIJkIJ .\E\left[ \mK\kro \mK \right]$ (see Proposition~\ref{lemma:psi2}). After some computations (a detailed proof is available in Appendix~\ref{anx:kempekro}), we have that
\begin{equation}\label{eq:projs}
\mR =\frac{1}{4} \mIJ\kro \mIJ   +  \frac{N-1}{4N}  \vv\vv^\mathrm{T}
\end{equation}
with $\vv = (1/\sqrt{N-1})  \left(\vu - (1/N) \vOne_{N^2}\right)$ and $\vu = \sum_{i=1}^N e_i \kro e_i$.

Consequently, $\mR$ is a linear combination of two following orthogonal projections:
\begin{itemize}
\item the first projection, generated by  $\mIJ\kro \mIJ$, is of rank $N^2-2N+1$, 
\item the second projection, generated by  $\vv\vv^\mathrm{T}$, is of rank $1$. 
\end{itemize}

As $\mIJ\kro \mIJ$ and  $\vv\vv^\mathrm{T}$ are orthogonal projections, the vector space $\mathds{R}^{N^2}$ (on which the matrix $\mR$ is operating) can be decomposed into a direct sum of four subspaces: 
\begin{itemize}
\item $\mathcal{S}_0 = \mathcal{I}m(\vv\vv^\mathrm{T}) \cap \mathcal{K}er \mIJkIJ $
\item $\mathcal{S}_1 = \mathcal{I}m(\vv\vv^\mathrm{T}) \cap \mathcal{I}m \mIJkIJ $
\item $\mathcal{S}_2 = \mathcal{K}er(\vv\vv^\mathrm{T}) \cap \mathcal{I}m \mIJkIJ $
\item $\mathcal{S}_3 = \mathcal{K}er(\vv\vv^\mathrm{T}) \cap \mathcal{K}er \mIJkIJ $
\end{itemize}
  
 As $\mIJkIJ \vv =  \vv$ (see Appendix \ref{anx:kempekro}), we have  $\mathcal{S}_0  = \{0\}$. 

Moreover, according to Eq.~(\ref{eq:projs}), we obtain that
\begin{eqnarray*}
 \mR\vx =  \left\{
\begin{array}{ll}
\left( \frac{1}{2} - \frac{1}{4N} \right) \vx & ~~~~ \forall \vx\in \mathcal{S}_1 \\
\frac{1}{4} \vx & ~~~~ \forall \vx\in \mathcal{S}_2 \\
0  & ~~~~ \forall \vx\in \mathcal{S}_3
\end{array}
\right.
\end{eqnarray*}
As a consequence, the non-null eigenvalues of $\mR$ are $1/4$ and $(1/2-1/(4N))$ which implies that $\rho\left(\mR \right) = 1/2 - 1/(4N)$. Hence, the convergence speed bound obtained by our general alternative approach  developed in this paper provides the true convergence speed for the Kempe's algorithm \cite{Kempe2003}.

%**************************************************************************%
\subsection{Comparison with B\'en\'ezit's algorithm}\label{sec:bene}

In \cite{Benezit2010a}, it has been shown that doing a multi-hop communication between sensors provides significant performance gain.
However, the proposed algorithm relied on a single-variate algorithm. In order to ensure the convergence of this algorithm, the double-stochasticity of the matrix update is necessary which implies a feedback along the route. The feedback can suffer from link failure (due to high mobility in wireless networks). To counter-act this issue, B\'en\'ezit proposes to get rid of the feedback by using  the Sum-Weight approach
\cite{Benezit2010}. In this paper, the authors established a general  convergence theorem close to ours. In contrast, they did not provide any result about convergence speed. It is worth noting that our convergence speed results can apply to the B\'en\'ezit's algorithm.

%**************************************************************************%
\subsection{Comparison with the single-variate  algorithms}\label{sec:mono}

If the following additional assumption holds,
\begin{description}
\item[{\bf(A3)}] The matrices of  $\mathcal{K}$ are column-stochastic,
\end{description}
one can easily show that all the weights $\vw(t)$ remain constant and equal to  $\vOne$, i.e., 
\begin{eqnarray*}
\forall t>0, ~& & \vw(t)^\mathrm{T} = \vw(0)^\mathrm{T}  \mP(t) = \vOne^\mathrm{T} \mP(t) = \vOne^\mathrm{T} \\
\textrm{and } ~& & \vx(t)=\vs(t)=\mK(t)^\mathrm{T} \vx(t-1).
\end{eqnarray*}

Therefore, the single-variate  algorithms (\cite{Tahbaz-Salehi2008}) with double-stochastic update matrices such as the \emph{Random Gossip} \cite{Boyd2004,Boyd2006}, the \emph{Geographic Gossip} \cite{Dimakis2008} can surprisingly be cast into the Sum-Weight framework. Moreover as $\Psi_1(t) = \|\vx(0)\|_2^2$ because all the weights stay equal to $1$, the proposed results about $\Psi_2(t)$ (that is Section~\ref{sec:psi2}) can be applied directly to the squared error for these algorithms. 

Let us re-interpret the work of Boyd {\it et al.} \cite{Boyd2006} (especially their section 2)  in the light of our results. In  \cite{Boyd2006}, it is stated that under doubly-stochastic update matrices $\mK(t)$,  the mean squared error at time $t$  is dominated  by $\rho\left(\E[\mK\mK^\mathrm{T}] - (1/N)\vOne\vOne^\mathrm{T} \right)^t $ and converges to $0$ when $t$ goes to infinity if 
\begin{equation}
\label{eq:boyd}
\rho\left(\E[\mK] - \frac{1}{N}\vOne\vOne^\mathrm{T} \right) < 1.
\end{equation}
Since $\mK(t)$ is doubly-stochastic, one can remark that $ \mIJ \mathbb{E}\left[\mathbf{K} \mathbf{K}^{T} \right] \mIJ= \mathbb{E}\left[\mathbf{K} \mathbf{K}^{T} \right]-(1/N)  \vOne\vOne^\mathrm{T}$. By following the approach developed in the footnote  of Section~\ref{sec:psi2}, we obtained directly the domination proven in \cite{Boyd2006}. Moreover, the condition corresponding to Eq.~(\ref{eq:boyd}) actually implies Assumption {\bf(B)}. Indeed, due to Eq.~(\ref{eq:boyd}) and the double-stochasticity of $\mK(t)$, one can remark that the maximum eigenvalue of $\E[\mK]$ is unique and equal to 1. Consequently, $\E[\mK]$ is primitive, and thus Assumption {\bf(B)} holds (see Lemma \ref{lemma:eqB}). Furthermore, in \cite{Boyd2006} (see section II-B) , it is stated that the condition corresponding to Eq.~(\ref{eq:boyd}) is only a sufficient condition and that the necessary and sufficient condition is the following one 
\begin{equation}
\label{eq:boyd2}
\rho\left(\E[\mK\kro\mK] - \frac{1}{N}\vOne_{N^2}\vOne_{N^2}^\mathrm{T} \right) < 1
\end{equation}
which is exactly the same expression as that in Lemmas~\ref{lem:Rbis}~and~\ref{lemma:rho}\footnote{Indeed, as the vector $\vv$ used in our formulation can be replaced with the left eigenvector corresponding to the eigenvalue $1$ (see the proof of Lemma~\ref{lemma:rho} for more details) which is proportional to $\vOne$ here due to the double-stochasticity of the update matrices}. Along with the reasoning detailed in Section~\ref{sec:resconv}, these two lemmas prove that under assumptions {\bf(A1)} and {\bf(A2)}, the condition corresponding to Eq.~(\ref{eq:boyd2}) is eventually necessary and sufficient when assumption {\bf(A3)} is also satisfied.  

\begin{figure}[ht]
\centering
\includegraphics[width = 0.9\columnwidth]{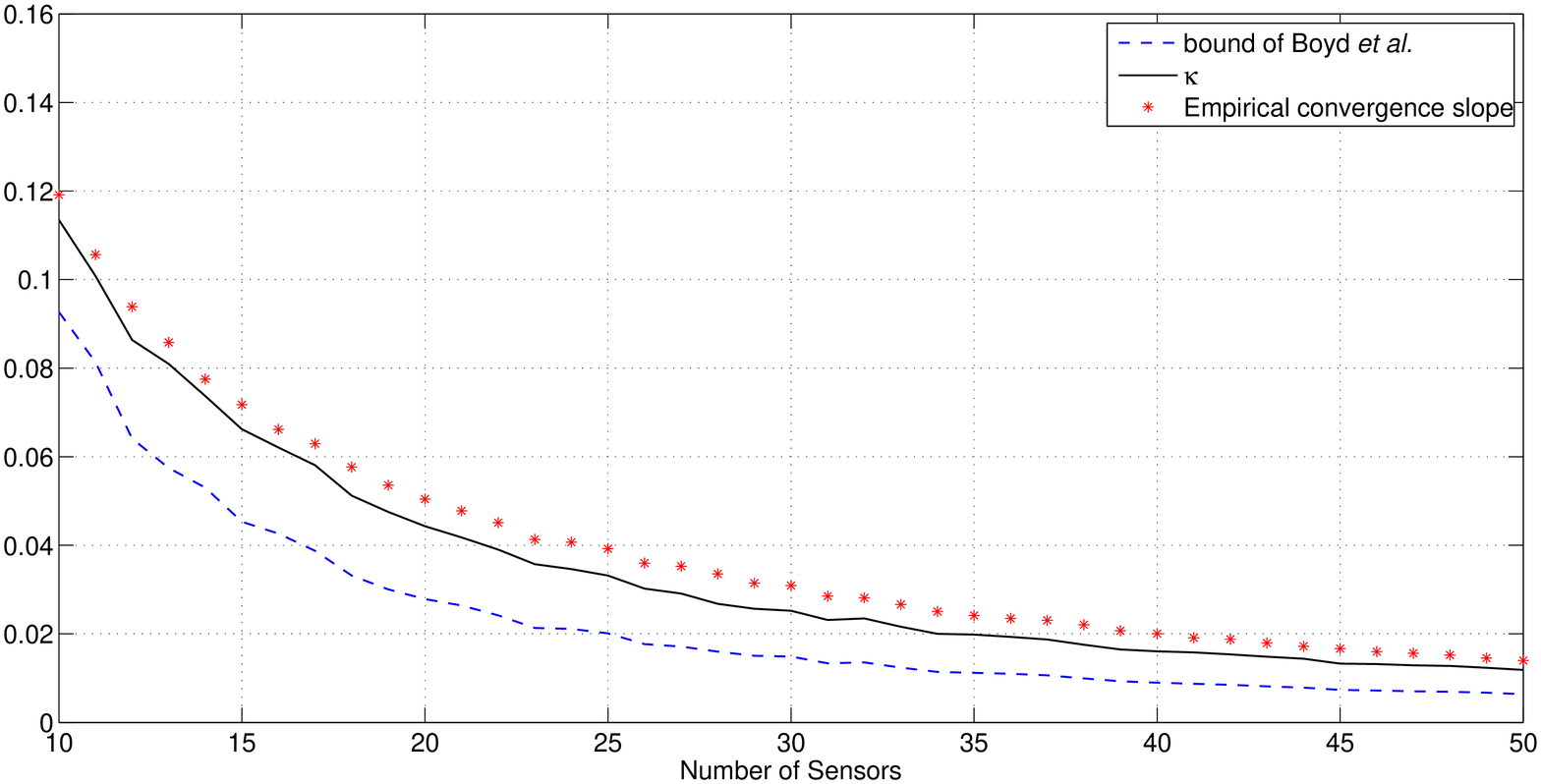}
\caption{\label{fig:slopeboyd} Empirical convergence slope of the \emph{Random Gossip},  the associated lower bound $\kappa$, and the bound given in \cite{Boyd2006}. }
\end{figure}

Moreover, according to Eq.~(19) (in \cite{Boyd2006}) and Eq.~(\ref{eq:nec}) (in our paper), we know that the mean squared error at time $t$  is upper bounded by $-\kappa't$ with $\kappa'=-\log(\rho\left(\E[\mK\kro\mK] - (1/N)\vOne_{N^2}\vOne_{N^2}^\mathrm{T} \right))>0$. However, as stated in Proposition~\ref{lemma:psi2}, the logarithm of the squared error scales with $- \kappa t$. Though these two spectral radii are less  $1$ and so ensure the convergence,  $\rho\left(  \mIJ \kro \mIJ. \E\left[\mK \kro \mK \right] \right)$ (\ie $\e^{-\kappa}$) exhibited in our paper is in general smaller than $\rho\left(\E[\mK\kro\mK] - (1/N)\vOne_{N^2}\vOne_{N^2}^\mathrm{T} \right)$ (\ie $\e^{-\kappa'}$) introduced in \cite{Boyd2006}. Hence, thanks to our approach, a tighter convergence speed bound has been derived. Numerical illustrations related to this statement are displayed on Fig.~\ref{fig:slopeboyd}.

%**************************************************************************%
%******************* Numerical illustrations ******************************%
%**************************************************************************%
\section{Numerical results}\label{sec:num}

In order to investigate the performance of distributed averaging algorithms over Wireless Sensor Networks, the use of Random Geometric Graphs (RGG) is commonly advocated. These graphs consist in uniformly placing $N$ points  in the unit square (representing the vertices of the future graph) then connecting those which are closer than a predefined distance $r$. A choice of $r$ of the form $\sqrt{r_0 \log(N)/N }$ with $r_0 \in [1,..,10]$ ensures connectedness with high probability when $N$ becomes large and avoids complete graphs (see \cite{Penrose2003} for more details). 

\begin{figure}[ht]
\centering
\includegraphics[width = 0.9\columnwidth]{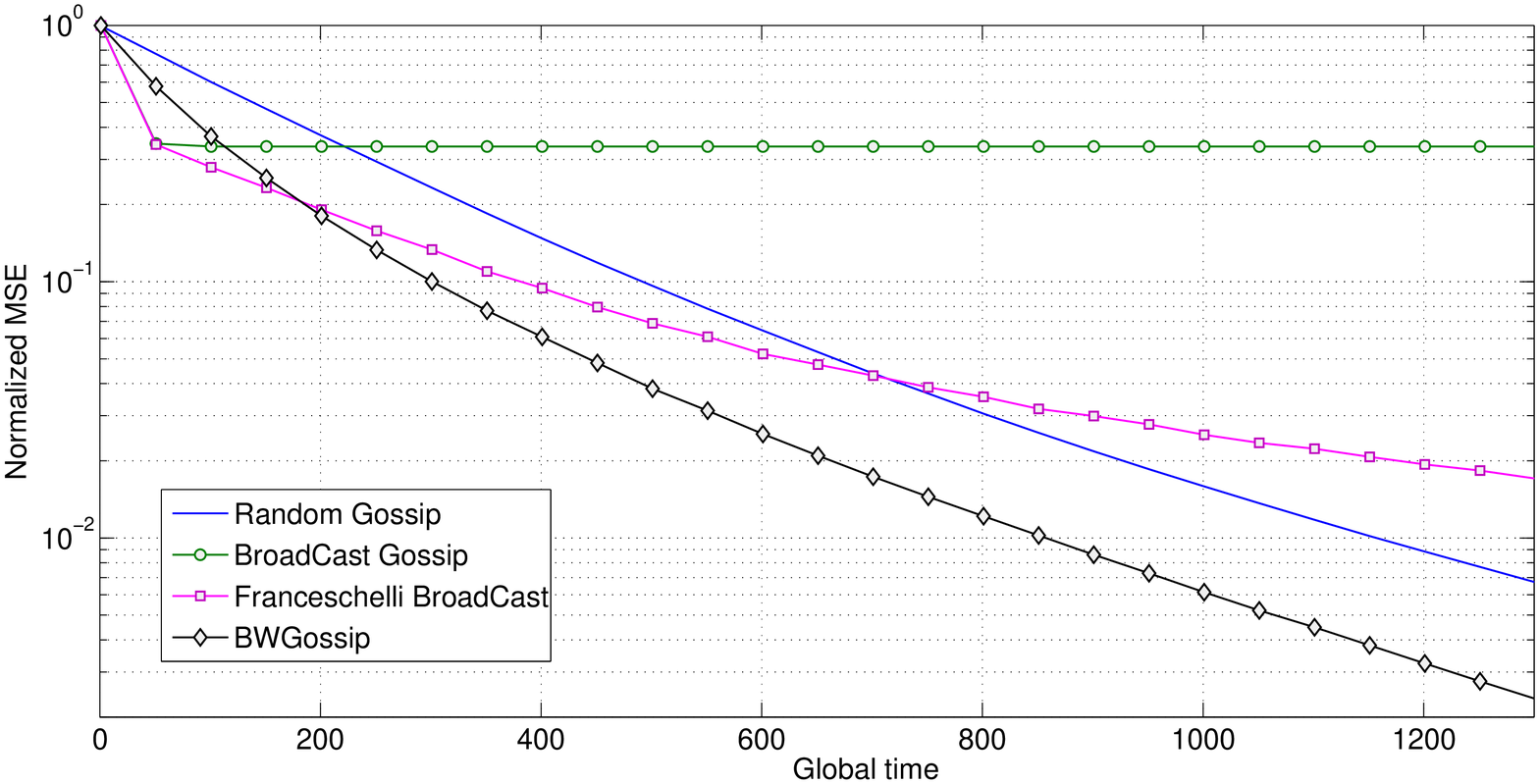}
\caption{\label{fig:algos} Mean squared error of the \emph{BWGossip} and other famous algorithms versus time.}
\end{figure}

In Fig.~\ref{fig:algos}, we plot the empirical mean squared error versus time for different gossip algorithms: i) the \emph{Random Gossip} \cite{Boyd2004} which is the reference algorithm in the literature; ii) the \emph{Broadcast Gossip} introduced in \cite{Aysal2009} which uses the broadcasting abilities of the wireless channel but does not converge to the average; iii) the algorithm introduced by Franceschelli in \cite{Franceschelli2011} which uses a bivariate scheme and seems to converge (no convergence proof is provided in the paper); and iv) the proposed \emph{BWGossip} algorithm.  A Random Geometric Graphs with $N=100$ sensors and $r_0=4$ has been considered. We remark that the \emph{BWGossip} algorithm outperforms the existing algorithms without adding routing or any other kind of complexity.

In Fig.~\ref{fig:clock}, we plot the empirical mean squared error for the \emph{BWGossip} algorithm versus time with different clock tuning coefficients (see \ref{sec:clock} and Eq.~(\ref{eq:lambda}) for more details). Compared to the algorithm without clock management ($\alpha=1$),  the convergence is much faster at the beginning with $\alpha=0$ but the asymptotic rate is lower; with  $\alpha=0.5$, the performance is better than the \emph{BWGossip} for any time.

\begin{figure}[ht]
\centering
\includegraphics[width = 0.9\columnwidth]{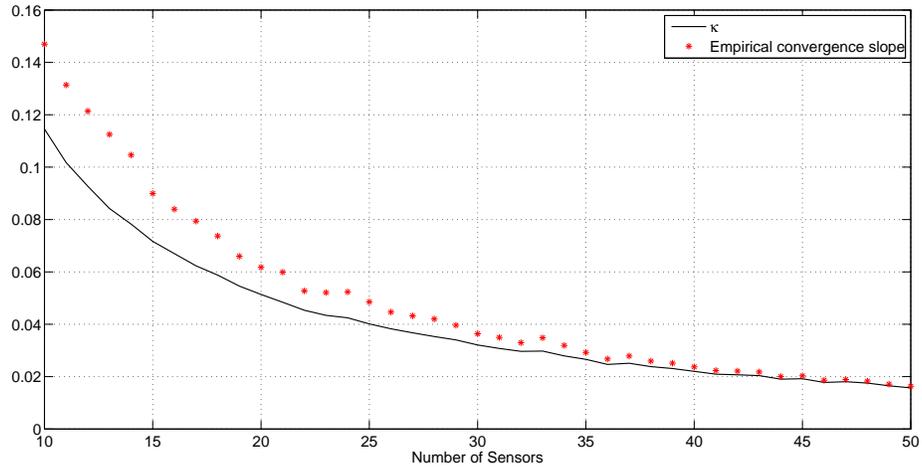}
\caption{\label{fig:slope} Empirical convergence slope of the \emph{BWGossip} and the associated lower bound $\kappa$.}
\end{figure}

In Fig.~\ref{fig:slope}, we display the empirical convergence slope\footnote{this slope has been obtained by linear regression on the logarithm of the empirical mean squared error. This regression makes sense since, for inspected algorithms, the mean squared error in log scale is almost linear for $t$ large enough as seen in Fig.~\ref{fig:algos}. } and the associated lower-bound $\kappa$ derived in Theorem~\ref{theo:speed} for the \emph{BWGossip} algorithm versus the number of sensors $N$. Different Random Geometric Graphs with $r_0=4$ have been considered. We observe a very good agreement between the empirical slope and the proposed lower bound. Consequently, our bound is very tight.

In Fig.~\ref{fig:slopeboyd}, we display the empirical convergence slope, the associated lower-bound $\kappa$, and the bound given in \cite{Boyd2006} for the {\it Random Gossip} algorithm versus the number of sensors $N$. The proposed bound $\kappa$ fits much better than the one proposed in \cite{Boyd2006}. Actually, the proposed bound matches very well the empirical slope (see Section~\ref{sec:mono} for more details).

\begin{figure}[ht]
\centering
\subfloat[\label{fig:MSEpe} Mean squared error versus time for different link failure probabilities.  ]{\includegraphics[width = 0.9\columnwidth]{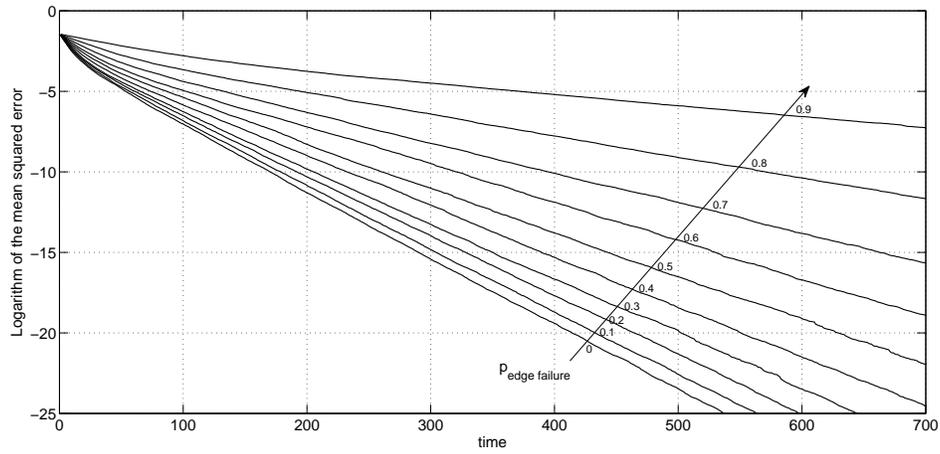}}\\
\subfloat[\label{fig:slopepe} Empirical convergence slope and the associated lower bound $\kappa$ versus link failure probabilities.  ]{\includegraphics[width = 0.9\columnwidth]{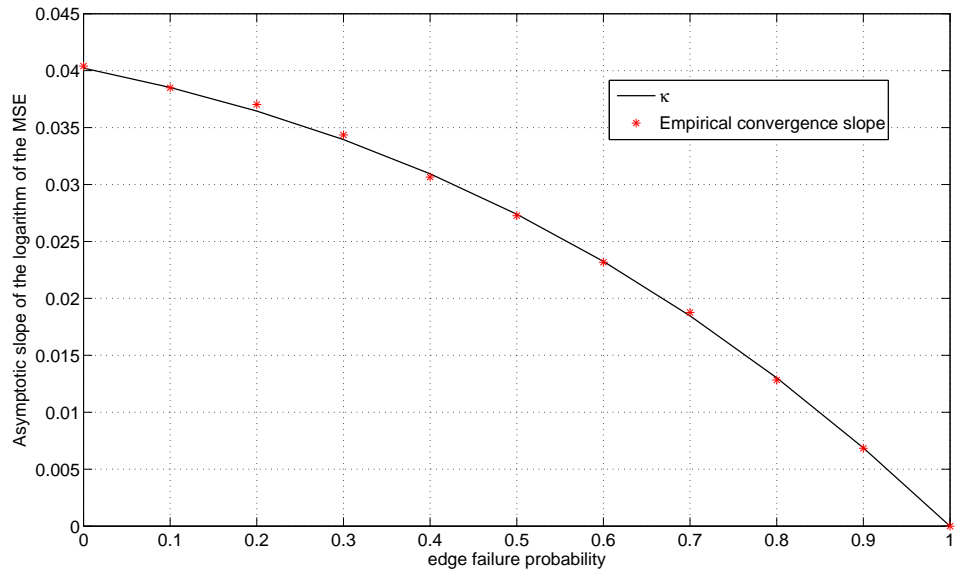}}
\caption{\emph{BWGossip} analysis in the presence of link failures.}
\label{fig:pe}
\end{figure}

Thanks to Fig.~\ref{fig:pe}, we inspect the influence of link failures in the underlying communication graph on the \emph{BWGossip} algorithm. We consider a Random Geographic Graph with $10$ sensors and $r_0=1$ onto which i.i.d. link failure events appear with probability $p_e$. In Fig.~\ref{fig:MSEpe}, we plot the empirical mean squared error of the \emph{BWGossip} versus time for different values of the edge failure probability $p_e$. As expected, we observe that the higher $p_e$ the slower the convergence but the MSE still exponentially decreases. 
Then, in Fig.~\ref{fig:slopepe}, we plot the empirical convergence slope and the associated bound $\kappa$ for different link failure probabilities. Here, $\kappa$ is computed according to a modified matrix set taking into account the link failures through different update matrices. We remark a very good fitting between our lower bound and the simulated results. Consequently, computing $\kappa$ on the matrix set including the link failures enables us to predict very well the convergence speed in this context.

%**************************************************************************%
%******************* Ccl **************************************************%
%**************************************************************************%
\section{Conclusion}\label{sec:cl}
In this paper, we have analyzed the convergence of the Sum-Weight-like algorithms (relying on two variables rather than one) for distributed averaging in a Wireless Sensor Network. We especially give a very precise insight on the convergence speed of the squared error for such algorithms. In addition, we proposed a particular Sum-Weight-like algorithm taking full advantage of the broadcast nature of the wireless channel. We observed that this algorithm significantly outperforms the existing ones.

%**************************************************************************%
%******************* Appendices *******************************************%
%**************************************************************************%
\appendices

%**************************************************************************%
%******************* Appendix A *******************************************%
%**************************************************************************%
\section{Proof of Lemma~\ref{lemma:eqB}}
\label{anx:eq}
{\bf {\bf (B)}  $\Rightarrow$ {\bf (B1)}}   Let denote by $\mK^{(u,v)}$ a matrix of $\mathcal{K}$ whose $(u,v)$-th coefficient is positive. As the graph associated with $\E\left[ \mK\right]$ is connected, then for all couples of nodes $(i,j)$, there is a path of finite length $L_{ij}<N$ from $i$ to $j$: $(i=u_1,..,u_{L_{ij}}=j)$. Consequently, the matrix $\mK^{i\to j} = \mK^{(u_1,u_2)} \mK^{(u_2,u_3)}..\mK^{(u_{L_{ij}-1},u_{L_{ij}})}$ verifies: $(\mK^{i\to j})_{i,j}>0$ which gives us a realization of $\mP(L_{ij})$ verifying  $(\mP(L_{ij}))_{i,j}>0$.\\
{\bf {\bf (B1)}  $\Rightarrow$ {\bf (B2)}} Let us take $L=\sum_{i,j=1}^N L_{ij} < 2 N^2$. Since each matrix has a positive diagonal according to Assumption~{\bf (A2)} then $\prod_{i,j=1}^N \mK^{i\to j}$ is a possible realization of $\mP(L)$ of strictly positive probability which is a positive matrix.\\
{\bf {\bf (B2)}  $\Rightarrow$  {\bf (B3)}}
If there is a $L<2N^2$  and a realization $\mathbf{p}$ of $\mP(L)$ so that $\Prob[\mP(L) =\mathbf{p}]>0$ and  $\mathbf{p}>0$, then $\mathbf{p}\kro\mathbf{p}$ is also positive. Since $(\mA\kro\mB).(\mC\kro\mD) = (\mA\mC) \kro (\mB\mD)$  for any matrices $\mA,\mB,\mC,\mD$ with the appropriate dimensions,  
\begin{eqnarray*}
\left(\E\left[ \mK\kro \mK\right]\right)^L &=& \left(\sum_{i=1}^M p_i \mK_i \kro \mK_i\right)^L \geq \Prob[\mP(L) =\mathbf{p}] . \mathbf{p}\kro\mathbf{p} > 0.
\end{eqnarray*}
Hence, $\E[\mK\kro\mK]$ is a primitive matrix.\\
{\bf  {\bf (B3)}  $\Rightarrow$  {\bf (B)}} First, we will calculate $\E\left[ \mK\right]\kro\E\left[ \mK\right]$ with respect to $\E\left[ \mK \kro \mK \right]$. So, 
\begin{eqnarray*}
\E\left[ \mK\right]\kro\E\left[ \mK\right] &=&  \sum_{i=1}^M \sum_{j=1}^M p_i p_j \mK_i \kro \mK_j \\
& \geq & \sum_{i=1}^M p_i^2 \mK_i \kro \mK_i \\
&\geq& (\min_j \ p_j) . \sum_{i=1}^M p_i \mK_i \kro \mK_i = (\min_j \ p_j) . \E\left[ \mK \kro \mK \right]
\end{eqnarray*}
Hence as it exists $k$ such that $ \left(\E\left[ \mK \kro \mK \right]\right)^k >0$, then  $\left(\E\left[\mK\right]\right)^k > 0$ so the primitivity of $\E\left[\mK\right]$ is proven.

\section{Derivations related to Section~\ref{sec:ana}}\label{anx:der}

\subsection{Derivations for Eq.~(\ref{eq:speed_kempe})}\label{anx:kempeL2}
According to Eq.~(\ref{eq:kk}), we have easily that 
\begin{eqnarray*}
\mK(t)\mK(t)^\mathrm{T} 
&=&  \frac{1}{4} \mI  + \frac{1}{4} \sum_{i=1}^N e_i e_{j_i(t)}^\mathrm{T}   + \frac{1}{4} \sum_{i=1}^N  e_{j_i(t)} e_i^\mathrm{T}  + \frac{1}{4} \sum_{i=1}^N \sum_{i'=1}^N e_i e_{j_i(t)}^\mathrm{T} e_{j_{i'}(t)} e_{i'}^\mathrm{T} 
\end{eqnarray*}
By remarking that $e_j^\mathrm{T}e_j=1$, we have 
\begin{eqnarray*}
\mK(t)\mK(t)^\mathrm{T} 
&=&  \frac{1}{2} \mI  + \frac{1}{4} \sum_{i=1}^N e_i e_{j_i(t)}^\mathrm{T}   + \frac{1}{4} \sum_{i=1}^N  e_{j_i(t)} e_i^\mathrm{T}  +  \frac{1}{4} \sum_{i=1}^N \sum_{\underset{i'\neq i}{i'=1}}^N e_ie_{j_i(t)}^\mathrm{T} e_{j_{i'}(t)}e_{i'}^\mathrm{T} 
\end{eqnarray*}

The randomness in $\mK(t)\mK(t)^\mathrm{T}$ is only due to the choice of the nodes $j_i(t)$ for $i=\{1,\cdots, N\}$. Therefore, each $j_i(t)$ will be modeled by a random variable $\ell(i)$ (independent of $t$). The random variables $\{\ell(i)\}_{i=1,\cdots, N}$ are i.i.d. 
and are uniformly distributed over $\{1,\cdots, N\}$. As a consequence, we obtain 
\begin{equation*}
\E[\mK\mK^\mathrm{T}]=  \frac{1}{2} \mI  + \frac{1}{4}    \sum_{i=1}^N e_i   \left(\frac{1}{N}\sum_{k=1}^N e_{k}^\mathrm{T} \right)  + \frac{1}{4}   \sum_{i=1}^N  \left(\frac{1}{N}  \sum_{k=1}^N  e_{k} \right) e_i^\mathrm{T}   + \frac{1}{4} \sum_{i=1}^N \sum_{\underset{i'\neq i}{i'=1}}^N e_i  \left(  \frac{1}{N^2} \sum_{k,k'=1}^N e_k^\mathrm{T} e_{k'} \right)        e_{i'}^\mathrm{T} \\
\end{equation*}
By remarking that $e_k^\mathrm{T}e_{k'}=0$ as soon as $k\neq k'$, we have 
$\sum_{k,k'=1}^N e_k^\mathrm{T} e_{k'}=N$. Furthermore,
\begin{eqnarray*}
\textrm{as } ~~ \sum_{k=1}^N  e_{k} =\vOne   &\textrm{and}& \sum_{i=1}^N \sum_{\underset{i'\neq i}{i'=1}}^N e_i e_{i'}^\mathrm{T}= \vOne\vOne^\mathrm{T}-\mI \\
\textrm{we obtain } ~~ \E[\mK\mK^\mathrm{T}] &=& \left( \frac{1}{2}  -  \frac{1}{4N} \right)\mI + \frac{3}{4} \mJ 
\end{eqnarray*}
It is then straightforward to obtain Eq.~(\ref{eq:speed_kempe}).  

\subsection{Derivations for Eq.~(\ref{eq:projs})}\label{anx:kempekro}
Once again, according to Eq.~(\ref{eq:kk}), we have easily that 
\begin{eqnarray}
\nonumber \mK(t)\kro \mK(t) &=& \frac{1}{4} \mI\kro \mI  + \frac{1}{4} \left( \sum_{i=1}^N e_i e_{j_i(t)}^\mathrm{T} \right) \kro \mI + \frac{1}{4} \mI \kro \left( \sum_{i=1}^N e_i e_{j_i(t)}^\mathrm{T} \right) \\
&+&  \frac{1}{4} \underbrace{\left( \sum_{i=1}^N e_i e_{j_i(t)}^\mathrm{T} \right) \kro \left(  \sum_{i'=1}^N e_{i'} e_{j_{i'}(t)}^\mathrm{T} \right)}_{\xi} \label{eq:def_xi}
\end{eqnarray}
Using the same technique as in Appendix \ref{anx:kempeL2} , we obtain that 
\begin{equation}\label{eq:jj}
\mathbb{E}\left[ \sum_{i=1}^N e_i e_{j_i(t)}^\mathrm{T} \right] = \sum_{i=1}^N e_i \left(\frac{1}{N}  \sum_{k=1}^N  e_{k}\right)= \mJ
\end{equation}
Thus, it just remains to evaluate $\mathbb{E}[\xi]$.  Let us first remark that 
$$\xi= \sum_{i=1}^N\sum_{\underset{i'\neq i}{i'=1}}^N    e_i e_{j_i(t)}^\mathrm{T} \kro e_{i'} e_{j_{i'}(t)}^\mathrm{T}   + \sum_{i=1}^N e_i e_{j_i(t)}^\mathrm{T} \kro e_{i} e_{j_{i}(t)}^\mathrm{T} $$
As a consequence, we have 
\begin{eqnarray*}
\mathbb{E}[\xi] &=& \frac{1}{N^2} \sum_{i=1}^N  \sum_{\underset{i'\neq i}{i'=1}}^N  \sum_{k=1}^N  \sum_{k'=1}^N  e_i e_{k}^\mathrm{T} \kro  e_{i'} e_{k'}^\mathrm{T}
+ \frac{1}{N} \sum_{i=1}^N  \sum_{k=1}^N  e_i e_{k}^\mathrm{T} \kro   e_{i} e_{k}^\mathrm{T} \\
&=& \frac{1}{N^2} \sum_{i=1}^N  \sum_{i'=1}^N  \sum_{k=1}^N  \sum_{k'=1}^N  e_i e_{k}^\mathrm{T} \kro  e_{i'} e_{k'}^\mathrm{T}   + \frac{1}{N} \sum_{i=1}^N  \sum_{k=1}^N  e_i e_{k}^\mathrm{T} \kro   e_{i} e_{k}^\mathrm{T} - \frac{1}{N^2} \sum_{i=1}^N   \sum_{k=1}^N  \sum_{k'=1}^N  e_i e_{k}^\mathrm{T} \kro  e_{i} e_{k'}^\mathrm{T} \\
\end{eqnarray*}
Using the well-known result on Kronecker product ( $(\mA\mB)\kro(\mC\mD)= (\mA\kro \mC)(\mB\kro \mD)$ for four matrices $\mA$, $\mB$, $\mC$, and $\mD$ with appropriate sizes), we have 
\begin{equation}\label{eq:exi}
\mathbb{E}[\xi] =  \mJ\kro\mJ  +  \frac{1}{N} \vu \vu^\mathrm{T} - \frac{1}{N^2} \vu \vOne_{N^2}^\mathrm{T}.
\end{equation}
Putting Eqs.~(\ref{eq:jj})-(\ref{eq:exi}) into Eq.~(\ref{eq:def_xi}), we get
\begin{equation*}
\E\left[ \mK\kro \mK \right] =\frac{1}{4} \mI\kro \mI + \frac{1}{4} \mJ \kro \mI + \frac{1}{4} \mI \kro  \mJ  + \frac{1}{4}  \mJ\kro\mJ +  \frac{1}{4N} \vu \vu^\mathrm{T} - \frac{1}{4N^2} \vu \vOne_{N^2}^\mathrm{T}.
\end{equation*}

Before going further, let us remark that
\begin{eqnarray}
\nonumber \mIJkIJ \vu &=& \sum_{i=1}^N (e_i-\frac{1}{N} \vOne \vOne^\mathrm{T} e_i) \kro (e_i-\frac{1}{N}\vOne \vOne^\mathrm{T} e_i)\\
\nonumber &=& \sum_{i=1}^N e_i \kro e_i - \sum_{i=1}^N (e_i  \kro \frac{1}{N} \vOne ) - \sum_{i=1}^N ( \frac{1}{N} \vOne \kro e_i ) +  \frac{1}{N^2} \sum_{i=1}^N  \vOne \kro \vOne \\
&=& \vu - \frac{1}{N} \vOne_{N^2}.\label{eq:vv} 
\end{eqnarray}
As a consequence, we have
\begin{eqnarray*}
\mR &=& \mIJkIJ .\E\left[ \mK\kro \mK \right] \\
 &=& \frac{1}{4} \mIJ\kro \mIJ   + \frac{1}{4N}  \left(\vu - \frac{1}{N} \vOne_{N^2}\right) \vu^\mathrm{T} - \frac{1}{4N^2}  \left(\vu - \frac{1}{N} \vOne_{N^2}\right) \vOne_{N^2}^\mathrm{T} \\
&=&  \frac{1}{4} \mIJ\kro \mIJ  +  \frac{1}{4N}  \vu\vu^\mathrm{T} -  \frac{1}{4N^2}  \vOne_{N^2} \vu^\mathrm{T}  - \frac{1}{4N^2} \vu  \vOne_{N^2}^\mathrm{T} + \frac{1}{4N} \mJ\kro\mJ
\end{eqnarray*}
Let us remind $\vv = \frac{1}{\sqrt{N-1}}  \left(\vu - \frac{1}{N} \vOne_{N^2}\right)$.
Thanks to Eq.~(\ref{eq:vv}), we have 
$$ \vv\vv^\mathrm{T} =  \frac{1}{N-1}  \left( \vu\vu^\mathrm{T} - \frac{1}{N}  \vOne_{N^2}\vu^\mathrm{T} - \frac{1}{N} \vu \vOne_{N^2}^\mathrm{T} + \mJ\kro \mJ \right)$$ which straightforwardly leads  to Eq.~(\ref{eq:projs}). 

In addition, note that using Eq.~(\ref{eq:vv}), we have $\mIJkIJ \vv =\vv$.

\newpage
$ $
\vspace{\stretch{1}} 

\vspace{\stretch{1}} 

\vspace{\stretch{1}} 

\vspace{\stretch{1}}

\vspace{\stretch{1}} 

\end{document}